%% file: main.tex
\documentclass[sigconf,anonymous=FALSE]{acmart}
\AtBeginDocument{%
  \providecommand\BibTeX{{%
    \normalfont B\kern-0.5em{\scshape i\kern-0.25em b}\kern-0.8em\TeX}}}
    
\setcopyright{acmcopyright}
\copyrightyear{2018}
\acmYear{2018}
\acmDOI{XXXXXXX.XXXXXXX}

\acmPrice{15.00}
\acmISBN{978-1-4503-XXXX-X/18/06}

\usepackage[commandnameprefix=ifneeded,todonotes={textsize=scriptsize}]{changes}
\usepackage{xspace}
\usepackage{listings}
\usepackage{libertine}
\usepackage{graphicx}
\usepackage{textcomp}
\usepackage{tikz}
\usepackage{caption}
\usepackage{subfig}

\usepackage{xcolor}
\usepackage{comment}
\usepackage{changes}
\usepackage{algorithm}
\usepackage{algorithmicx, algpseudocode}
\usepackage{multirow}
\usepackage{tablefootnote}
\usepackage{threeparttable}
\usepackage{makecell}
\usepackage{array}
\usepackage{enumitem}
\usepackage{color}
\usepackage{soul}
\usepackage{booktabs}
\usepackage{float}

\definechangesauthor[name={Hyoung}, color=red]{hs}

\newcommand{\sysname}{{CookieCruncher}\xspace}

\begin{document}

\title{Crumbled Cookies: Exploring E-commerce Websites' Cookie Policies with Data Protection Regulations}

\author{Nivedita Singh}
\email{singhnivvy@g.skku.edu}
\orcid{0000-0001-6225-3669}
\affiliation{%
  \institution{Sungkyunkwan University}
  \streetaddress{Seobu-ro, Jangan-gu}
  \country{South Korea}
  \postcode{2066}
}

\author{Yejin Do}
\email{dyj001213@g.skku.edu}
\affiliation{%
  \institution{Sungkyunkwan University}
  \streetaddress{Seobu-ro, Jangan-gu}
  \country{South Korea}}

\author{Yongsang Yu}
\email{eysl14198@g.skku.edu}
\orcid{0009-0004-6790-6488}
\affiliation{%
  \institution{Sungkyunkwan University}
  \streetaddress{Seobu-ro, Jangan-gu}
  \country{South Korea}
}
\author{Imane Fouad}
\email{imane.fouad@inria.fr}
\affiliation{%
  \institution{Inria, Univ Lille}
  \country{France}
}
\author{Jungrae Kim}
\email{dale40@skku.edu}
\affiliation{%
 \institution{Sungkyunkwan University}
 \streetaddress{Seobu-ro, Jangan-gu}
 \country{South Korea}}

\author{Hyoungshick Kim}
\email{hyoung@skku.edu}
\authornotemark[1]
\affiliation{%
  \institution{Sungkyunkwan University}
  \streetaddress{Seobu-ro, Jangan-gu}
   \country{South Korea}}

\begin{abstract}
  Despite stringent data protection regulations such as the General Data Protection Regulation (GDPR), the California Consumer Privacy Act (CCPA), and other country-specific regulations, many websites continue to use cookies to track user activities. Recent studies have revealed several data protection violations, resulting in significant penalties, especially for multinational corporations. Motivated by the question of why these data protection violations continue to occur despite strong data protection regulations, we examined 360 popular e-commerce websites in multiple countries to analyze whether they comply with regulations to protect user privacy from a cookie perspective. We were interested in whether the cookie policies of websites differ from country to country and whether the strictness of data protection rules in a country affects the cookie policies of websites in that country. Our investigation explored cookie attributes and their links to potential security and privacy breaches, such as cross-site scripting (XSS) and cross-site request forgery (CSRF). We found that approximately 84\% of all examined cookies are susceptible to XSS attacks. While preventive mechanisms against CSRF attacks lie at the mercy of browsers, our results are concerning, revealing that 81\% of cookies are prone to these attacks. We also discovered masquerading cookies, where third-party cookies pose as first-party cookies, making it difficult to distinguish them from first-party cookies. These deceptive cookies can track users without their knowledge or consent. To our knowledge, we are the first study to analyze how well e-commerce websites in different countries uphold the GDPR/CCPA and country-specific regulations. Our findings suggest that many websites still use insecure cookie policies. Therefore, there is an urgent need for uniform and consistent cookie policies. This need also underscores the importance of enacting stringent regulations that define the attributes associated with cookies.
\end{abstract}



\settopmatter{printfolios=true}
\maketitle

\section{Introduction}
Cookies are an intrinsic element of web applications, initially designed to streamline the state between client and server, but their use has expanded as application needs have evolved. While some cookies are essential for the proper functioning of the website, a majority of them are used for user tracking and advertisement~\cite{bollinger2022automating}. Recently, web user tracking has garnered substantial attention from the public due to its proliferation in data privacy incidents. These cookies are under scrutiny because of their covert acquisition of extensive user information, subsequently monetized by the ad tech sectors, often leading to notable incidents such as the Cambridge Analytica scandal~\cite{Julia}. 

In 2019, it was reported that about 90\% of all the websites use tracking cookies~\cite{solomos2019clash}. As a result, they have become a focal point in ongoing legislative endeavors. Concerns over user privacy arising from the mismanagement of cookies have been voiced over the years in several press media outlets, research publications, and prominent academic conferences~\cite{CookiesTerror}. Consequently, a more comprehensive examination of this topic has emerged, encompassing diverse perspectives~\cite{EUcookieLaw}. This exploration delves into technical considerations related to cookie syncing~\cite{fouad2018missed} among web entities, as well as legislative measures governing user privacy and behavior.

Third-party sources inject JavaScript code to enhance user experiences and ensure websites run smoothly. Nonetheless, this cross-domain inclusion of third-party JS code invites privacy and security risks making them an essential nuisance~\cite{chen2021cookie}. In an online web context, third-party tracking and breaches are unarguably the most persistent intrusion in users' data. Though several efforts have been made to restrict these third-party cookies tracking, not much attention has been paid to how first-party cookies are also engaged in user tracking and other breaches~\cite{180596}.


To restrict these violations, the European Union introduced the General Data Protection Regulation (GDPR),~\cite{GDPR1}, which enforces strict guidelines on the collection and retention of personal data, requiring companies to obtain informed consent. Similarly, the California Consumer Privacy Act (CCPA)~\cite{CCPA} provides consumers with more control over the personal information that businesses collect. Recognizing the urgent need for data protection rules, several countries are racing to implement/revise their own data protection regulations, such as PIPA for South Korea~\cite{PIPA}, while others have yet to do so, such as the PDP bill, India~\cite{IndiaDPBill}. 

 To our interest, we selected various e-commerce platforms because they are among the most visited websites in the world, and unfortunately, they carry security weaknesses~\cite{pagey2023all}. Businesses benefit from ad-tech organizations such as Google Analytics, which often contain blocklists such as Criteo. However, these gains often come at the price of potential personal data breaches affecting consumers. To make matters worse, some major players (e.g., Amazon and AliExpress) offer services in various parts of the globe but often fail to adhere to country-specific data protection rules. This non-compliance has resulted in regulatory authorities inflicting significant fines on many leading organizations for enabling advertisement cookies even before users choose their consent~\cite{AmazonFine, FineOnAmazonGDPR1}. These players create sophisticated tracking mechanisms in collaboration with major ad-tech companies by enabling cross-site data transfer~\cite{fouad2022my}. Our findings revealed that their pervasive cookies constantly monitor user choices and preferences even when these major players operate within specific countries. This observation highlights the extensive influence of these cookies, which transcends geographical boundaries.
We are motivated to explore, despite the existence of strict data protection rules, how platforms find loopholes to infringe. To this end, we explore various cookie attributes contributing to security breaches and privacy violations. We raise the following research questions:
 
\begin{itemize}
 \item RQ1: Do e-commerce platforms comply with GDPR, CCPA, and other privacy legislations?

 \item RQ2: How vulnerable are user profiles on shopping websites? If so, to what extent?

 \item RQ3: To what extent do cookies violate the rules?
\end{itemize}

\begin{figure}[h]
  \centering
  \includegraphics[width=\linewidth]{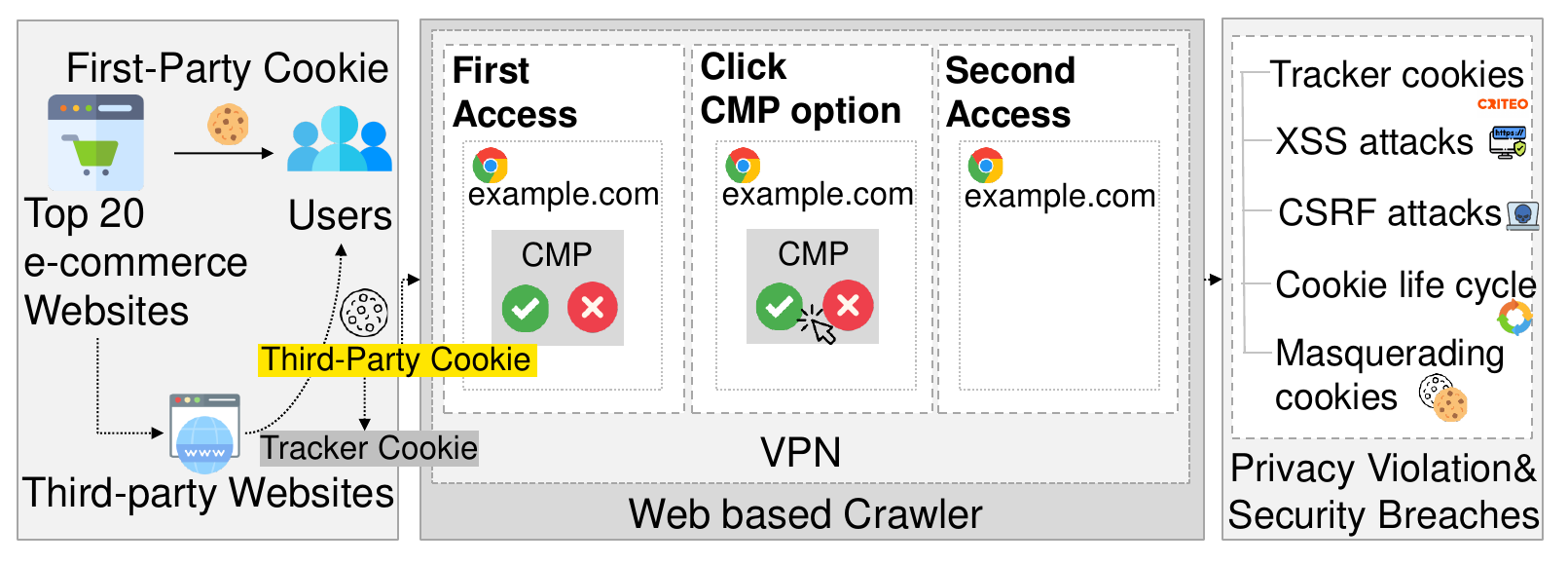}
  \caption{Overview of \sysname.}
  \Description{Overview of work}
  \label{fig:Overview}
\end{figure}

To address these research questions, we developed a tool called \sysname. Figure~\ref{fig:Overview} shows an overview of \sysname, which takes an e-commerce website as input, searches for a consent management platform (CMP) if one exists, waits for user consent, collects cookies, and then analyzes their behavior. \sysname used this process to crawl 11,223 cookies from real-world websites and assess their attributes. Our analysis highlights how third-party cookies can lead to potential breaches.

\textbf{Our contributions} 
First, \sysname collected and analyzed 11,223 cookies from e-commerce platforms, finding potential security and privacy violations of 84\% and 81\% against XSS and CSRF attacks, respectively. Of the 11,000 cookies, 37\% are third-party cookies, of which 67\% are tracking cookies. For example, we found that nearly 65\% of the top 20 third-party cookies from 18 countries were ad trackers, and some of them were even blocklisted cookies (e.g., Criteo, Adnxs).

\sysname also evaluated that approximately 40\% of cookies transcend their lifecycle, which could help construct comprehensive user profiles to unveil behaviors and interests. While earlier investigations focused on identifying violations and user tracking cases, \sysname emphasizes uncovering the underlying causes of these infringements. We believe that addressing violations demands more than just identifying the means; it necessitates a thorough examination of their characteristics, primarily referred to as cookie attributes. For example, \sysname evaluated how third-party cookies track users in the first party's name by hiding their true identity. We discovered these types of cookies as masquerading cookies.
Our findings could prompt legislators to rethink cookie policies, with an emphasis on their process of creation, sharing, and dissemination across websites.

\section{Background}
Before we dive into how various cookies can be abused, we first provide basic information about cookies~\cite{carmi2017cookies}.

\subsection{Cookies}
Cookies are diminutive text files strategically embedded in web browsers to maintain user states during browsing sessions, enabling websites
to store and manage user-specific information across multiple sessions. Figure~\ref{fig:Attribute description} shows the attributes of an example cookie. Common cookie attributes include:
\begin{figure}[h]
  \centering
  \begin{lstlisting}[
    basicstyle=\footnotesize, %or \small or \footnotesize etc.
]
{
    "domain": ".360yield.com",
    "expires": 1697000572.987183,
    "httpOnly": false,
    "name": "tuuid_lu",
    "path": "/",
    "priority": "Medium",
    "sameParty": false,
    "sameSite": "None",
    "secure": true,
    "session": false,
    "size": 18,
    "sourcePort": 443,
    "sourceScheme": "Secure",
    "value": "1689224572"
}
\end{lstlisting}
  \caption{Third-party cookie on Amazon.}
  \Description{Third-party cookie on Amazon (Top shopping website of the USA).}
  \label{fig:Attribute description}
\end{figure}

\begin{enumerate}
    \item \textbf{domain}: This attribute specifies the domain for which the cookie is valid. In Figure~\ref{fig:Attribute description}, a 360yield.com (third-party) cookie is placed on the website www.amazon.com.
    \item \textbf{expires}: Represents cookie lifecycle in Unix timestamp.
    \item \textbf{httpOnly}: This means the cookie is exclusively accessible through the HTTP or HTTPS protocol; setting it to "true" limits access to HTTP(S) requests, preventing client-side script access..
    \item \textbf{name}: Identify the purpose of the cookie.
    \item \textbf{path}: This attribute specifies the URL path within the domain for which a particular cookie is valid. In  Figure~\ref{fig:Attribute description}, `/' specifies that for cookie `tuuid\_lu,' all pages of the domain are valid.
    \item \textbf{priority}: Determines the priority.
    \item \textbf{sameParty}: Determines whether the cookie belongs to the domain, the user visit, or a different domain. In Figure~\ref{fig:Attribute description}, it is set as `false' as the particular cookie belongs to third party.
    \item \textbf{sameSite}: This attribute, controls or prevents the cookie when it is sent with cross-origin requests. This attribute is responsible for preventing cross-site forgery attacks. In Figure~\ref{fig:Attribute description}, it is set as `None,' which means it can be sent on cross-origin requests.
    \item \textbf{secure}: This attribute defines whether the cookie is only transmitted over secure HTTPS connections or not.
    \item \textbf{session}: Categorising cookies into session or persistent.
    \item \textbf{size}: This attribute represents the size of the cookie.
    \item \textbf{sourcePort}: This attribute specifies the source port from which the cookie was created and shared.
    \item \textbf{sourceScheme}: This attribute indicates the source scheme used to set the cookie.
    \item \textbf{value}: This attribute in the cookie determines the specific content or data associated with the cookie. In  Figure~\ref{fig:Attribute description}, it is given in the form of a unique identifier or code.
\end{enumerate}

These cookies enable ad-tech companies to trace users' cross-site behaviors adeptly, rooted in their visited sites and inferred interests. Subsequently, this insight fuels the delivery of tailor-made advertisements, enriching the user experience. In a recent study, it has been reported that ~91\% of an average users' browsing history could be traced successfully ~\cite{bashir2018diffusion}. This highly intimidating revelation prompts a compelling need for technical discourse and immediate interventions.

\subsection{Legislation for cookies and data privacy}
In this tangled scenario, various regulatory bodies have crafted strict legislation to create a cookie ecosystem that preserves and respects user choices. Recital 30 of GDPR mentions cookies as online identifiers for profiling and identification~\cite{GDPR}.
Likewise, within the California region of the USA, the California Consumer Privacy Act (CCPA)~\cite{CCPA} 
elevates privacy rights and safeguards for residents of California, mandating that businesses furnish consumers with notifications regarding their privacy protocols. These data protection rules force companies and businesses to obtain consent before processing cookies. 
It is worth noting that CCPA is a state law for the California region but is applicable outside of California. For the USA, there is a proposal for federal law `American Data Privacy Protection Act (ADPPA)' 2023 ~\cite{American_Data_Privacy_and_Protection_Act, American_Data_Privacy_and_Protection_Act(Forbes)}. 

In line with GDPR, most countries formulated their own data protection rules to safeguard users' privacy. For instance, PIPA for the Republic of Korea~\cite{PIPA}, New Zealand Privacy Act 2020~\cite{Newzealandact}, Brazil-General data protection (LGPD) 2020~\cite{BrazilAct}, etc.
Nevertheless, many countries are addressing user privacy by building upon existing regulations and creating new ones. For instance, in India, user protection is anchored in the Information Technology Act of 2000, while a fresh data protection bill is being formulated~\cite{IndiaDPBill}. Conversely, countries like Australia, governed by the Australian Federal Privacy Act of 1988, have introduced a novel data protection law focused on consumers known as the Consumer Data Right (CDR)~\cite{Consumer_Data_Right, Australia_Data_Protection_Legislation}.

\section{Methodology}
\sysname generated a dataset by crawling the top 20 revenue-generating e-commerce platforms globally from 18 different countries (360 websites), and analyzed 11,223 cookies in total. We chose high-revenue e-commerce websites in a specific country, maintaining consistency by selecting sites recommended by Statista (\url{www.statista.com}) and Similarweb (\url{www.similarweb.com}). This procedure has strengthened our belief in the sustained popularity and stability of these domains. With the intention of obtaining a global dataset for our research, we employed a virtual private network (VPN) to retrieve information directly from its source. To achieve this, we utilized Proton VPN (\url{www.protonvpn.com}) to establish a connection via a designated country's proxy server.

We implemented our cookie analysis framework on top of a custom crawler which is based on the open-source web browser Chromium (version 113.0.5672.126 (Official Build) (64-bit)). We used a server based on Ubuntu/VMware Virtual Platform, 2 CPU cores with 13th Gen Intel(R) Core(TM) i9-13900KF. The experiment was performed in July 2023.
\begin{figure}[h]
  \centering
  \includegraphics[width=\linewidth]{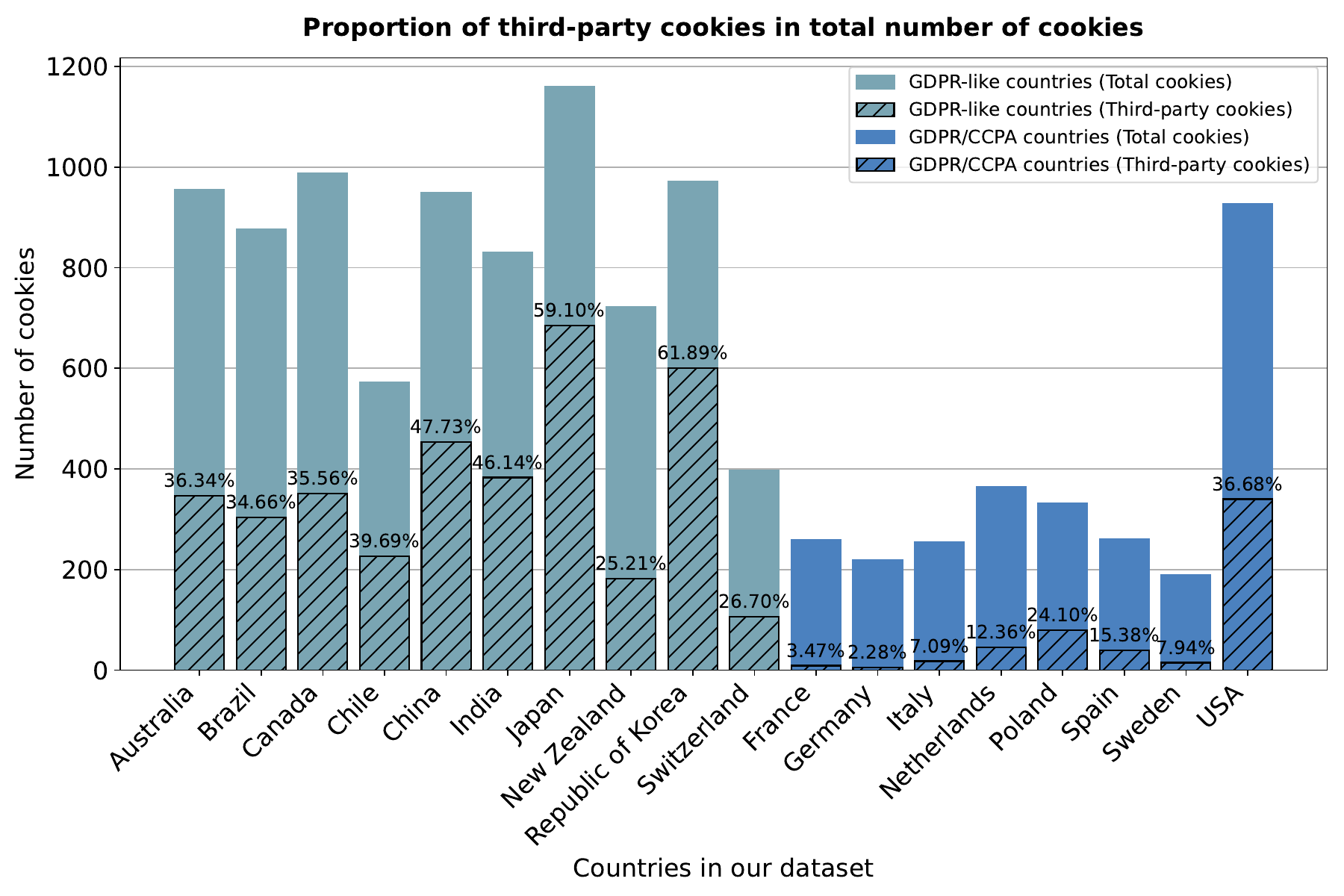}
  \caption{Third-party cookie occurrences for all 18 countries.}
  \Description{Top twenty-third party organizations.}
  \label{fig:cookies and Third-party}
\end{figure}
\subsection{Countries selection}
Although there may have been other bifurcations, for our studies, we have deliberately chosen 18 countries and categorized them into two groups based on their compliance with data protection regulations. The initial category consists of countries with more stringent rules, 7 European nations that adhere to GDPR and ePrivacy directive guidelines, and the USA (CCPA). Countries that make up our EU's list, are France, Germany, Italy, Netherlands, Poland, Spain, and Sweden. We specifically selected these EU countries based on the presence of maestro e-commerce platforms. Here the maestro organizations are those organizations that establish a presence globally such as Amazon, Alibaba, etc.
In the subsequent list, we have designated 10 additional countries, namely Australia, Brazil, Canada, Chile, India, Japan, New Zealand, the Republic of Korea, and Switzerland. The rationale behind selecting these particular countries is their enforcement of rigorous data protection regulations akin to GDPR or their active efforts toward implementing stringent data protection measures~\cite{kawintiranon2021towards, johnson2022economic}.

\subsection{Crawler and handling CMP platform}
\sysname have used a selenium-based crawler~\cite{jha2022internet} and collected the data after establishing a proxy connection to each country. According to the GDPR, CCPA, and other data protection rules, user data can only be collected after getting the consent of the user.
Regulations require websites to seek user consent before processing their data, which is why CMPs have become necessary.
During our experiment, \sysname discovered, that employment of CMP for GDPR-like countries was limited, while CMP was properly inbuilt for most of the GDPR/CCPA countries. For instance, Australia contains CMP for only two websites among 20, Bigw and Theiconic, while for France, all 20 platforms contain CMP except Amazon, carrefour, e.lecerc, Groupon and zalando.

For CMP selection, we intentionally click on the `Accept all' button to find the top third-party organization for a particular country, later used to study their behavior. Apparently, in GDPR-like countries, where many of the websites do not contain the CMP, we sincerely collect the cookies data and analyze it. Our crawler collects in three instances- the first being when it lands on any website, the second, cookies after clicking CMP (if any), and finally, the cookies on the second visit after clicking the consent button. We collected a total of 11k cookies from the visited 360 websites.

\subsection{Data analysis}
The crawler is fed with the aforementioned e-commerce list, loads the relative web pages, and the test browser collects all the cookies.
Next, the comprehensive examination of cookies commences by closely examining their overall characteristics. In the post-analysis, we discovered that there is a significant presence of third-party cookies. Our experimental and empirical data analysis revealed several interesting features; specifically, 90\% of our crawler generated were at the optimum risk. Our attributes-based analysis revealed that a good number of the third-party cookies appear to belong to giant ad tech and mostly blocklisted advertising organizations.

\section{Analysis of real-world websites}

To address RQ1, we use \sysname to examine the attributes of cookies, regardless of whether they are first-party or third-party, and based on the findings, we report the instances of potential security breaches and privacy violations. 

\subsection{Complex web of third-party and tracker cookies}

Cookies were not created with malicious intent. They were initially designed to help websites remember information about users, such as login credentials, user preferences, and items in a shopping cart. However, over time, cookies have been morphed into a tool for tracking users across the web, which has led to concerns about privacy and security~\cite{cahn2016empirical, hu2020tangled}. In the big-data era, third-party cookies are intentionally deployed by data brokerage firms, online ad agencies, and tracking applications (e.g., DoubleClick, 360yield, Criteo)~\cite{bollinger2021analyzing}. Third-party cookies can be used to track users' browsing behavior and later assist ad-tech agencies in creating user profiles, thereby challenging users' privacy.

We scrutinized the crawled third-party cookies and found that most of them were global trackers. The abundance of third-party cookies, including persistent cookies, poses challenges for users in tracking and controlling their online presence.

\subsubsection{Cookies pervasiveness}
For our analysis, \sysname filtered out the first-party cookie and third-party cookie by considering the cookie domain name (See Figure ~\ref{fig:cookies and Third-party}). We now detail the list of all 18 countries in Appendix~\ref{sec:Third-party information and cookie attributes}.

Though cookie's pervasiveness is evident in both the category of our chosen list, there is a mark difference in their proportions. The proportion of cookies in our GDPR-like countries list easily surpasses the GDPR/CCPA countries. Evidently, in our GDPR-like countries data set, the majority of the countries have third-party cookies proportion in the range of around 40-about 50\%, while it hardly reaches about 40\% maximum in the USA (strict category).
Though countries in our GDPR/CCPA category comply with uniform regulations, they show a variation among themselves. Particularly the cookies proportion of Germany is the least while Poland tops the list with the USA as an exception. The reduced third-party cookies proportion of Germany among EU countries is in line with the previous reports on an average number of trackers per website~\cite{jha2022internet}.
Less surprisingly, these trackers are in excess proportion in GDPR-like countries. Though our analysis presents that countries with GDPR/CCPA rules have better data protection, nonetheless, it does not ensure complete protection in safeguarding privacy. At the same time, there is much to do for our GDPR-like countries.
\subsubsection{Cookie before and after analysis}\sysname analyzed the cookie appearance behaviors for those websites that contain CMP. Understanding the factors that third-party cookies are often deployed to run the advertisement business, but in many instances user has no choice but to give his consent. For instance, in theiconic.com (Australia), 22 third-party cookies (including Adnxs, doubleclick. net, criteo,360yield, etc.) appeared before the user gave consent, while the other cookies either first-party or third-party, appeared after the user feed his consent to CMP. The issue of concern is whether these cookies are linked to the potential security breaches that we address in the following section.


\begin{figure}[h]
  \centering
  \includegraphics[width=\linewidth]{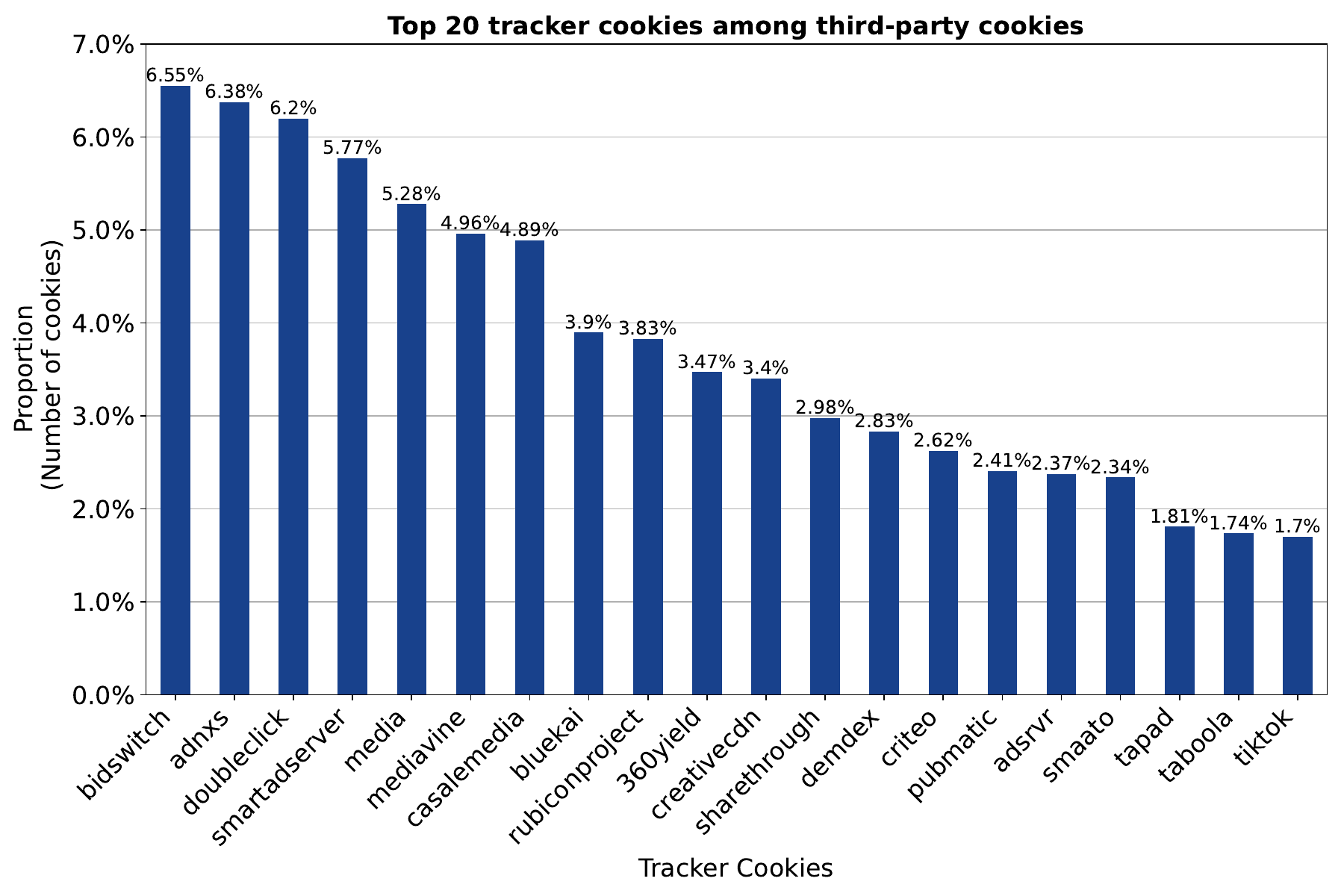}
  \caption{Top 20 tracker cookies from third-party list.}
  \Description{Top 20 third-party appearances from all 18 countries in our dataset.}
   \label{fig:trackerdomain}
\end{figure}
\subsubsection{Analysis by GDPR-like countries}
We now analyze the occurrence of third-party cookies and their variations among our GDPR-like countries list (See Figure~\ref{fig:fig4} (Appendix)). 
For these countries, out of total cookies, 43\%  are third-party cookies, consinting mostly trackers discussed further in the next section. 
This analysis showcases the name of some targeted ad agencies, particularly `Adnxs,' which redirects the browsers to add unwanted browser extensions, malicious software updates, adult sites, online web games, and other illegitimate content. It is highly dangerous that it is popularly called “Tracking the tracker's data”~\cite{Adnxs},~\cite{cahn2016s}.
Out of the top 20 third-party cookies, Bing occupies the top position with 5.06\% occurrence while adsrvr the lowest (1.54\%). In our in-depth analysis, we found that widerplanet is a predominant active tracker cookie in South Korean websites which is a targeting advertising service based on a big data source pool. 
\subsubsection{Analysis by GDPR/CCPA countries}
Third parties, which remain dominant in GDPR-like countries, prove their global presence by appearing in GDPR/CCPA countries as well, but with a slightly reduced proportion (19\%) (see Figure~\ref{fig:fig5} (Appendix)). 
For instance, Bing (occurrence 5.06\%) in the GDPR-like countries appears with somehow the same occurrence (5.07\%) in GDPR/CCPA countries.
 \subsubsection{Analysis by all  countries}
To keep an eye on the pervasiveness of the third-party cookies across all the countries, we generated a combined plot of all 18 countries considering 360 websites (see Figure~\ref{fig:Thirdpartyallcountry} (Appendix)). Our analysis revealed that third-party cookies more or less follow the same pattern globally. For instance, Adnxs, a concerning tracker, exhibited a presence of 4.36\% in GDPR-like scenarios, 4.36\% in GDPR/CCPA countries, and thus an overall occurrence of 4.46\% in the third-party tracker list. 
Among a total of 11k cookies, \sysname evaluated 37\% cookies are third-party cookies, and these cookies are accessible to all pages of specific platforms. In further investigation, we evaluated that 98\% of third-party cookies can trace every page by setting its path attribute as `/'
 \subsubsection{Tracker cookies, among third-party cookies}
 \sysname performed analysis on the third-party cookie list using filterlist~\cite{Filterlist} and filtered out cookies set by a tracker, resulting in 67\% of the total tracking cookies, to track user over different websites. bidswitch and Adnxs are the top two trackers in our dataset (see Figure~\ref{fig:trackerdomain}).\\
\textbf{Our recommendation}
- Enforcing rigorous policies to limit the usage of third-party tracker cookies and their associated attributes is imperative.

\subsection{Misconfiguration of third-party cookie attributes}
\begin{figure}[h]
  \centering
  \includegraphics[width=\linewidth]{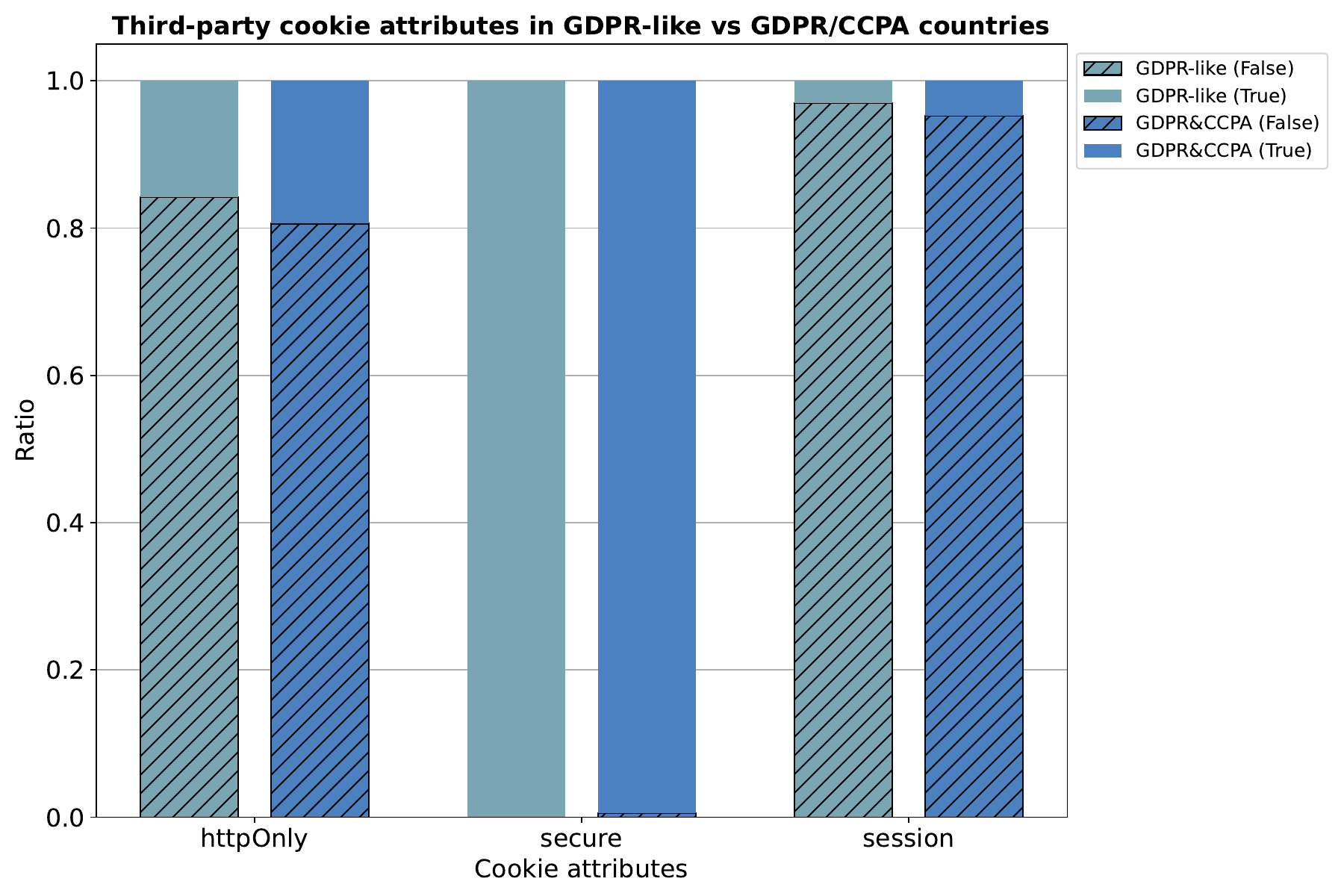}
  \caption{Third-party cookie attributes set to `true' or `false.'}
  \Description{Cookie attributes set to `true' or `false' and respective trends for all countries}
   \label{fig:XSS}
\end{figure}

\sysname investigated three cookie attributes from the third-party list, namely, `httpOnly,' `secure,' and `session.' Knowing the fact that for the proper functioning of a website, some flags are intentionally set as false. So to check the potential abuses, \sysname used only third party, including the tracker and advertisement cookies, to evaluate further. The flags of these cookies can be either set as `true' or `false,' which has a different implication on the security and privacy of the website and users. 
`httpOnly' attributes are considered as a defense in depth strategy ~\cite{zhou2010aren}. Enabling this flag as `true' acts as a safeguard against cross-site scripting (XSS) attacks. However, research indicates that while they don't completely eradicate XSS attacks, they do render them significantly more challenging by thwarting cookie theft. From Figure~\ref{fig:XSS}, it is evident that `httpOnly' flags are, in general, set to `false' in high proportion, approximately 84\% for all the cookies (RQ2 and RQ3). This situation can potentially attract intruders to inject malicious code that might initiate an XMLhttp request using the TRACE method, thereby possibly obtaining the `httpOnly' cookie in the message echoed back by the server~\cite{grossman2003cross}. 

The `secure' attribute setting to true can not prevent the possibility of an XSS attack as observed jointly by the httpOnly attribute (see Figure~\ref{fig:XSS}).

We took these three attributes into consideration as this combination to find the possibility of XSS attacks.
In summary, our analysis of these three attributes from all 18 countries shows their susceptibility to potential security attacks (RQ2).\\
\textbf{Our recommendation} - A combination of these three attributes must be judiciously handled to ensure user security.
\subsection{Invitation to CSRF attack}
Cross-site request forgery, popularly known as a one-click attack, is an exploit of a website where unauthorized actions are unintentionally submitted by a user resulting in financial loss or performing other sensitive operations. In 2016, the `sameSite' attribute of the cookie was introduced to prevent CSRF attacks to add an additional security layer as a defense mechanism.
 The `sameSite’ attribute for cookies has three possible flags. `Strict,' `Lax,' and `None.' 
`Strict' ensures the cookie is only sent with same-site requests, providing maximum security but potentially breaking some cross-site functionality. 
`Lax' allows the cookie to be sent with same-site and top-level navigation requests, striking a balance between security and compatibility. 
`None' allows the cookie to be sent with all requests, including cross-site, which is the very least secure. 
Cookie rules~\cite{CookieCSRF2019}proposed that browsers should treat the lack of the `sameSite' attribute as lax.  In a further refinement in 2020 ~\cite{CookieCSRF2020}, cookies with sameSite=None were additionally required to use the `secure' flag.
From our analysis (see Figure~\ref{fig:samesite} (Appendix)), we found that in the most dangerous CSRF attack, strict cookies' best armor has been least implemented, creating a welcoming door for this attack. Particularly in GDPR-like countries,
 this cookie proportion is lower compared to the GDPR/CCPA countries list. 
Our data aligns with a recent report showing that only a few websites safeguard their authentication cookies by enabling strict or lax ~\cite{compagna2021preliminary}. 
Though this SSC- a key defense against CSRF only came to the limelight when Chrome browser adopted it in 2020~\cite{Samesitelaw1, Samesitelaw4, PortSwigger}.
\begin{figure}[t]
  \centering
  \includegraphics[width=\linewidth]{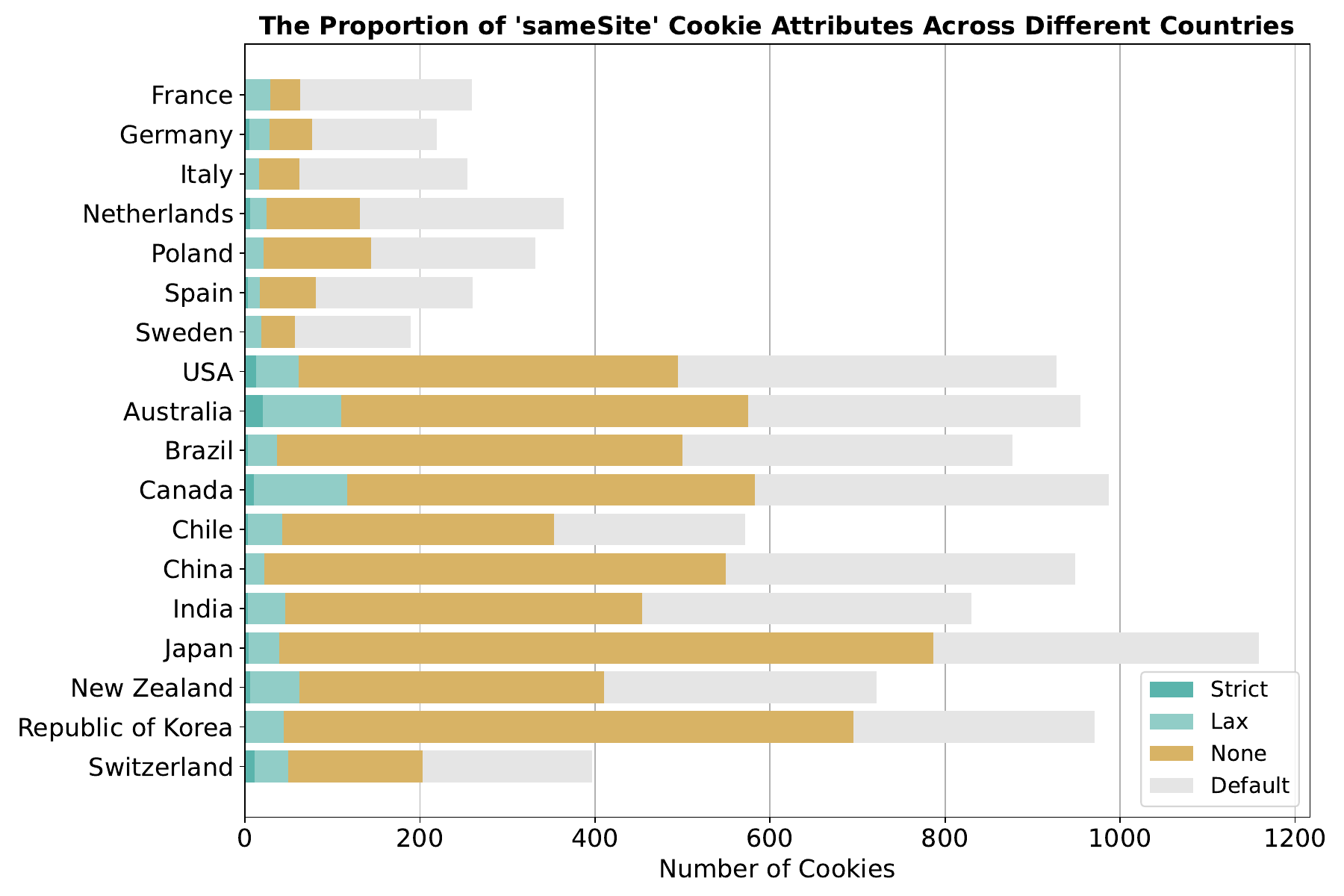}
  \caption{sameSite attribute for three lists (GDPR-like countries, GDPR/CCPA countries, and all 18 countries together).}
  \Description{sameSite attribute for three lists (GDPR-like countries, GDPR/CCPA countries, and all 18 countries together).}
   \label{fig:fig10}
\end{figure}
Moving to the other attribute, `None,' \sysname found a high proportion of cookies are set to none in GDPR-like countries compared to GDPR/CCPA countries (RQ2).
Initially, if the `sameSite' attribute was set as blank or with an invalid value (we defined it as default), the browser was compelled to handle such cookies as if they were tagged with sameSite=None. While this scenario was advantageous for websites minimally impacted by the implementation of `sameSite,' it resulted in the new security measure remaining dormant by default. However, in a follow-up draft in 2019, browsers were mandated to interpret the absence of `sameSite' attribute as sameSite=lax ~\cite{Samesitelaw3}. 
In another subsequent refinement in 2020, the browsers were prescribed to use the `secure' flag to the cookies configured with sameSite=None ~\cite{Samesitelaw3}. At this juncture, it is quite obvious that nations adhering to strict data protection regulations, similar to our category list of GDPR/CCPA countries, exhibit a slightly lower prevalence of cookies labeled with sameSite=None compared to their counterparts. Nevertheless, when considering a comprehensive analysis of all cookies across countries, a notable portion of cookies marked with sameSite=None could raise concerns about user privacy, warranting apprehension. Another feature that draws our attention is the maximum number of cookies tagged with sameSite=default.
Leaving the `sameSite' attribute as default places the burden on the browser to determine how the flag for this attribute is configured (see Figure ~\ref{fig:fig10}). However, it is important to highlight that if this attribute's flag is set to its default value, the corresponding cookie must also possess a specific `secure' field set to `true.'
Our examination reveals for potential security breach, underscoring the need for a reevaluation of this attribute (RQ2).

\begin{table}
  \caption{`sameSite' attribute with default flag (out of total cookie available for particular country's twenty e-commerce platforms) and respective `secure' attribute with `false' flag (FF).}
  \label{tab:freq1}
  \begin{tabular}{cccl}
    \toprule
    \textbf{Index}&\textbf{Country}&\textbf{Default}&\textbf{`secure' (FF)}\\
    \midrule
         1  &   Australia           &   39\%    &   86\%\\
         2  &   Brazil              &   42\%    &   91\%\\
         3  &   Canada              &   40\%    &   76\% \\
         4  &   Chile               &   38\%    &   95\%\\
         5  &   China               &   42\%    &   91\%\\
         6  &   India               &   45\%    &   86\%\\
         7  &   Japan               &   32\%    &   87\%\\
         8  &   New Zealand         &   43\%    &   84\%\\
         9  &   Republic of Korea   &   28\%    &   98\%\\
         10 &   Switzerland         &   48\%    &   70\%\\
         11 &   France              &   75\%    &   73\%\\
         12 &   Germany             &   64\%    &   69\%\\
         13 &   Italy               &   75\%    &   65\%\\
         14 &   Netherlands         &   64\%    &   69\%\\
         15 &   Poland              &   56\%    &   75\%\\
         16 &   Spain               &   68\%    &   71\%\\
         17 &   Sweden              &   69\%    &   67\%\\
         18 &   The USA             &   46\%    &   74\%\\
  \bottomrule
\end{tabular}
\end{table}

 Table~\ref{tab:freq1}  illustrates a disconcerting scenario. In the second column, which represents the default, we observe the total count of cookies lacking the `sameSite' attribute, thereby shifting the responsibility to browsers for defining this field. However, as elucidated earlier, in such instances, the `secure' field must be set to 'true'.

In the fourth column of the same table, we find a worrisome trend. Among the total cases under the default case, a certain percentage is marked as `false', giving rise to concern. For instance, considering France as an example, 75\% of cookies for that specific country are set as default, and within this subset, a substantial 73\% (RQ3) have their `secure' field flag set to `false'. Another inference from the same table could about Chile, where ~38\% cookies were set as default while 98\% of them carry a `false' flag for `secure' attribute. 
This pattern holds `true' for other countries as well. The situation is even more troubling for GDPR/CCPA countries. Despite their lower cookie count, the alignment with Figure~\ref{fig:samesite} (Appendix)) is evident. This figure indicates that in the case of GDPR/CCPA countries, a higher proportion of cookies are left to the browser's discretion, resulting in a significant number of these cookies having their `secure' field set as `false.' These cookies are susceptible to CSRF attacks. To find a security vulnerability in this case, \sysname considered all cookies as these abuses can be invited either first-party or third-party. When the user browses any e-commerce, even if he rejects all third-party cookies, still vulnerable to these attacks by first-party cookies. \\
\textbf{Our recommendation} - The `sameSite' attribute assumes a pivotal role in managing CSRF attacks, representing a significant milestone in deterring such perilous security breaches. Our investigation highlights persisting gaps in attribute implementation, underscoring the need for heightened legislative focus to address these intricacies effectively.
\begin{figure}[h]
  \centering
  \includegraphics[width=\linewidth]{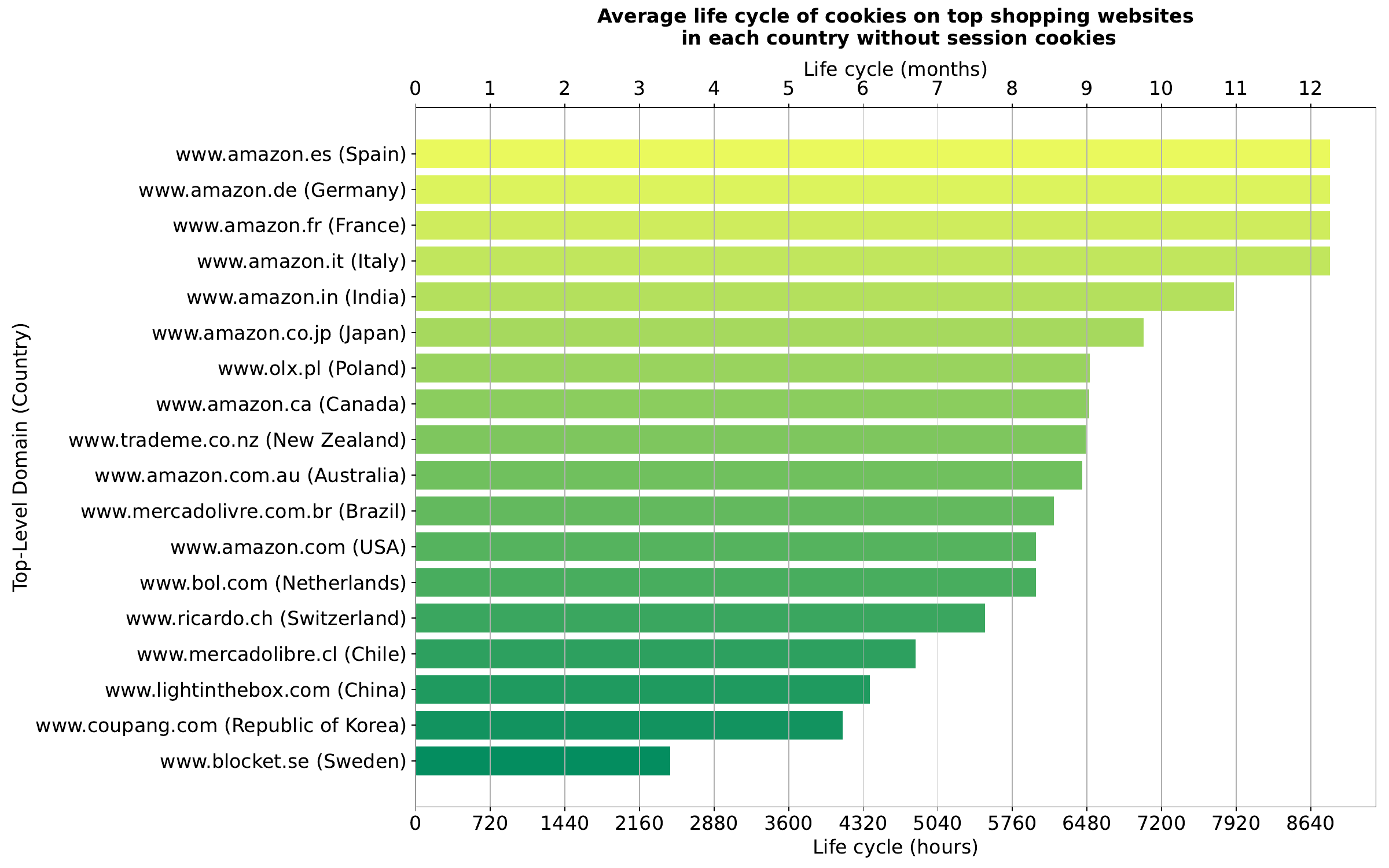}
  \caption{Average lifecycle of cookies on top visited e-commerce websites in each country.}
  \Description{Average lifecycle of cookies on top visited shopping websites in each country}
   \label{fig:fig12}
\end{figure}
\subsection{Cookies transcending their age}
According to the regulation, the websites declare the expiration time of personal information, which also applies to cookie~\cite{Cookie_Expiration_Time}.
It is crucial to verify the duration cookie.
You may gather the impression that our conversation revolves around none other than the lifecycle of cookies. A cookie containing user interest related information should be retained for a designated duration, which is why a cookie with a brief lifespan is regarded as more secure.
The eDirective Privacy specifies the cookie life span for not more than 12 months~\cite{GDPRCookieLifecycle}. This means the persistent cookies (the cookie that declares its lifespan) should not remain for more than 12 months. Our analysis results in Figure~\ref{fig:fig12} and (see Table~\ref{tab:table2}) raise concerns for users' data privacy. They depict many instances of violations regarding the cookie's lifespan.
In our analysis, we have chosen the average lifespan of cookies available from the prominent e-commerce/shopping websites without the session cookies.
As it is now well established that cookies store critical information like website visits, hobbies, interests, and purchasing trends.
Such humongous user data collection through cookies deployment leaves personal information vulnerable~\cite{pisano2022evolving}.
Despite this, several countries have paid less heed to cookie's lifespan
~\cite{Cookielifecycleforothercountries}, hence we took GDPR/ePrivacy Directive to compare the violations for all countries in our dataset.
However, several research reports strictly advocate on the lifespan of cookies must be established~\cite{naithani2022curtailing}.
Here we don’t bring any discussion on the nature of cookies simply to avoid ambiguity on the type of cookies to be essential, persistent, or strictly necessary. We straightforwardly focus on the cookies transcending their lifespan and their potential implications on data privacy. Before delving into further discussion, we highlight that to plot (see Figure~\ref{fig:fig12}), we have chosen the top revenue-generating e-commerce of each country in the category. On the very first preliminary inspection, it is clear that the majority of the countries violating the cookie's life span fall in our GDPR/CCPA countries category list such as Spain, Germany, Italy, and France (RQ1 and RQ2). It is quite contradictory that GDPR/CCPA countries are violating their own strict rules. This trend is not too far to guess, thanks to the maestro organization in these countries. Unfortunately, even after paying a huge fine in the past by maestro companies~\cite{FineOnAmazonGDPR}.
\begin{table}
  \caption{Cookie lifecycle (`Expire') potential violation}
  \label{tab:table2}
  \begin{tabular}{p{2cm} c p{2cm} c}
    \toprule
    \textbf{GDPR-like country} & \textbf{Violation} & \textbf{GDPR/CCPA country} & \textbf{Violation} \\
    \midrule
    Australia & 41\% & France & 36\% \\
    Brazil & 42\% & Germany & 32\% \\
    Canada & 44\% & Italy & 31\% \\
    Chile & 42\% & Netherlands & 32\% \\
    China & 40\% & Poland & 43\% \\
    India & 43\% & Spain & 36\% \\
    Japan & 52\% & Sweden & 38\% \\
    New Zealand & 43\% & The USA & 40\% \\
    South Korea & 38\% & -- & -- \\
    Switzerland & 42\% & -- & -- \\
    \bottomrule
  \end{tabular}
\end{table}
our results reveal a persistent trend. On the contrary, several countries in the GDPR-like countries category (their top-revenue e-commerce platform) perform better in terms of cookies' lifespan violation.

 Table~\ref{tab:table2} illustrates the number of cookies in each country that exceed one year rule for cookie lifespan. For instance, among the entirety of cookies present on twenty shopping websites in Australia, 41\% of these cookies are in breach of the rule, and this pattern continues for other countries as well.
Cookies lifespan violation is as little as 31.4\% and highest at 52.41\%  for Italy and Japan, respectively (see Table~\ref{tab:table2}).
It is essential to mention here that more the lifespan a cookie possesses, more vulnerable it will be. As a rule of thumb, these transcending cookies are more vigilant to user interest.\looseness=-1\\
\textbf{Our recommendation}
- Our investigation has unveiled instances where cookies exceeded a one-year duration, providing ample time for the compilation of comprehensive user profiles encompassing interests, behaviors, and decision-making patterns.
We earnestly appeal to policymakers to place a heightened emphasis on crafting stringent regulations to govern cookie lifespans precisely.
\begin{figure}[h]
  \centering
  \includegraphics[width=\linewidth]{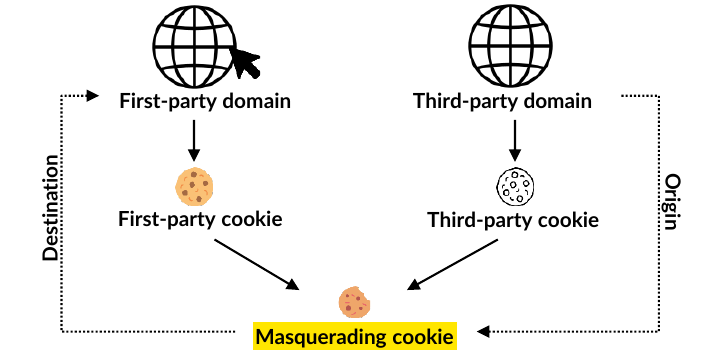}
  \captionsetup{width=0.8\textwidth}
  \caption{Overview of masquerading cookies.}
  \Description{Masquerading cookie example.}
   \label{fig:fig13}
\end{figure}
\subsection{Maestro websites and their masquerading cookies}
In this section, we elucidate how maestro websites employ their Masquerading cookie to weave a complex cookie ecosystem, tragically placing users’ privacy at risk.
While crawling, we landed up finding a third-party cookie behaving like a first-party and thus assisting in users' behavior tracking, to these cookies, we coined the name Masquerading Cookie. For example, while crawling to collect our cookies data from Shein.fr- a top revenue-generating company in France, we found cookies from Shein.com- a top Asian e-commerce organization (see Figure~\ref{fig:fig13}).
In order to authenticate our findings, we conducted website crawls with a `Reject all' approach, followed by subsequent data analysis. Our investigation uncovered a noteworthy discovery: despite the rejection of third-party cookies, deceptive masquerading cookies persist within the e-commerce platform. Afterward \sysname conducted the investigation of other attributes of the same cookies, and those were aligned with our findings, such as for these cookies the attribute `SameParty' carries a False flag ensuring these cookies do not belong to the same domain or first party.
Hence these masquerading cookies are conflicting third-party cookies camouflaged as first-party cookies.
The ubiquitous nature of maestro websites on the one hand benefits the user by their services very efficiently but at the stake of consumers' privacy. On the other hand, these maestro websites are the root origin of these masquerading cookies.
 It is obvious that these conflicting cookies may go unnoticed in many instances, making it challenging to distinguish between first-party and third-party cookies.
 (see table~\ref{tab:my-table} (Appendix)) Our empirical investigation revealed a significant occurrence of this phenomenon, particularly among globally serving maestro websites. \sysname answered our question that we initially introduced.

\textbf{Our recommendation} - Our findings expose the perilous possibility of third-party tracking under the guise of a first-party due to domain conflicts. This underscores a significant security concern. We hold the expectation that policymakers will thoroughly address this matter, reconsidering the precise delineation of first-party and third-party entities.
\section{Discussion and Limitations}
Our comprehensive examination of cookie attributes reveals a prevalent occurrence of intricate networks involving intermediaries or third parties on web pages. These cookies, originating from diverse domains, are subsequently employed for user tracking and security breaches.
Contemporary data protection regulations are earnestly striving to safeguard user privacy, despite the occasional presence of undesirable content on web pages. While one perspective reveals that the implementation of these rigorous data protection regulations will establish a barrier-like framework to curtail non-essential data exchange among third parties, our stance contends that a more nuanced approach, holds the potential to uncover the underlying causes behind these incidents.
Currently, a lack of established guidelines exists for defining individual cookie attributes. Each attribute of a cookie and its correlation with another attribute, such as the `sameSite' attribute requiring the `secure' attribute to be set as 'true,' holds significance in thwarting malicious attacks. These interconnected attributes with their respective flags weave a complex interconnected cookies network making it nearly impossible for users to evade the abuses. Therefore, a distinct legal framework is imperative to meticulously address every cookie attribute and its interrelations.
We believe our research suggests that lawmakers will likely institute more rigorous regulations to precisely delineate cookie attributes. This is essential, as cookies play a pivotal role in user tracking and monitoring online activities. Thus advocating for an urgent standard cookie policy.\\
Our work has a few limitations. Firstly, in identifying tracker cookies, we used a filterlist mechanism instead of more advanced methods. 
Secondly, in some of our studies, we adopted GDPR rules as a standard without analyzing nation-specific regulations, which we plan to explore in future research.
\section{Related work}
Cookies are among the simplest arsenals used in tracking mechanisms. 

Sanchez-Rola et al.~\cite{sanchez2021journey} defined a set of roles in the cookie ecosystem scenario related to cookie creation and sharing. This study highlights that cookies form a complex network of interconnections between organizations that could share the cookies at the end of a complex chain involving a middleman. In another instance, Sanchez-Rola et al. 
~\cite{sanchez2019can} analyzed EU websites and surprisingly found that cookie tracking remains ubiquitous. In this study, they revealed that though GDPR has impacted how data is being processed globally, nearly 90 percent of web visits remained traceable. 
Likweise, Bollinger et al. 
~\cite{bollinger2021analyzing},
through their studies, found many instances of GDPR violations. Fouad et al.
~\cite{fouad2018missed}
through their studies revealed that even the most popular tracking detection methods, such as EasyList and EasyPrivacy, miss a significant portion of cookie trackers, specifically failing to detect 25.22\% and 30.34\% of identified trackers, respectively. 
While these studies have demonstrated several occasions of violation of data protection rules, our study is the first of its type where we tried to analyze the cause of security and privacy violations for privacy regulation-owning countries.

To summarize, our research stands apart in two key ways. Initially, we identified the potential security and privacy breaches stemming from the presence of these cookies carrying undesired attributes and their linked flags. Secondly, we endeavored to delve into the intricate reasons behind the challenging nature of mitigating tracking cookies, primarily due to their conflicting identities.
\section{Conclusion}

Our studies revealed that the major culprit for cookie-related violations and breaches is cookie attributes rather than the cookies themselves, which have been unfairly blamed in the past. In this paper, we performed a comprehensive measurement campaign and highlighted possible violations and breaches beyond the GDPR and CCPA. Our in-depth analysis demonstrates the potential for global third parties to track users and expose their profiles to vulnerabilities. Our findings indicate that cookie attributes and settings of false or None could be the reasons for XSS and CSRF attacks. These vulnerabilities compromise cookie security and increase the risk of unauthorized access and data interception. These attacks can be even more risky if the cookies are long-lived or if they are masquerading cookies.

At this juncture, we urge policymakers to rethink the standardization of cookie policy, focusing on ways to improve the creation, sharing, and dissemination of cookies across websites.

In our future work, we aim to develop a tool to enhance security and privacy during web browsing. This tool will take a URL as input and strive to provide users with a URL devoid of tracking cookies. It will also evaluate and modify potentially harmful cookie attributes. The tool will first check for a cookie consent banner and consider the user's choice (acceptance or rejection). If no consent banner is found, the tool will proceed autonomously.
 \bibliographystyle{ACM-Reference-Format}
 \bibliography{References}
\appendix
\balance
\input{Appendix}
\end{document}

%% file: Appendix.tex

\appendix

\section{Third-party information and cookie attributes}
\label{sec:Third-party information and cookie attributes}

\begin{figure}[H]
  \centering
  \includegraphics[width=\linewidth]{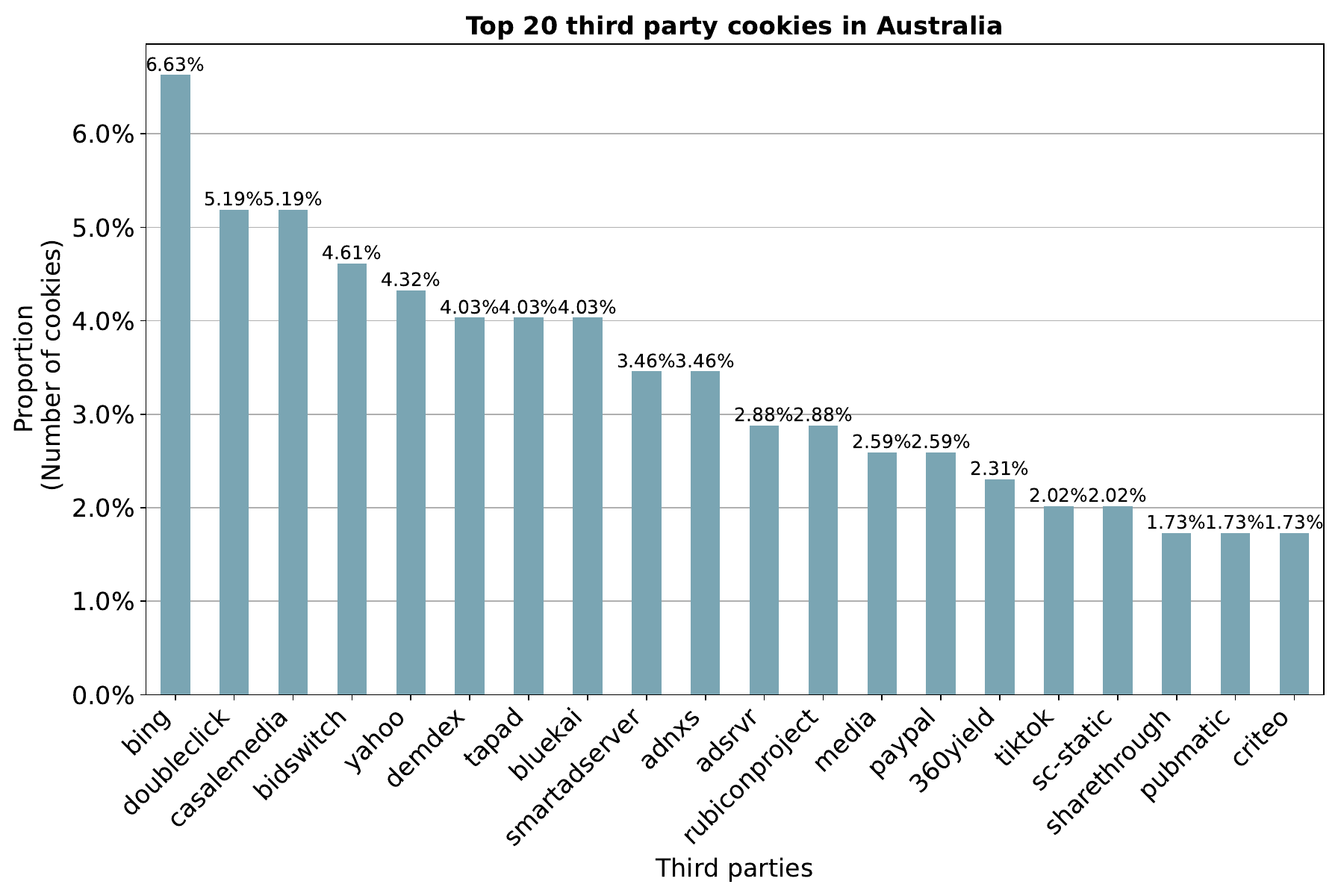}
  \caption{Third-party cookies distribution in Australia. On a global scale, the prominent trackers `bing' and `doubleclick' hold the top positions among trackers in Australia.}
  \Description{Third-party cookies distribution in Australia.}
  \label{fig:fig 14}
\end{figure}

\begin{figure}[H]
  \centering
  \includegraphics[width=\linewidth]{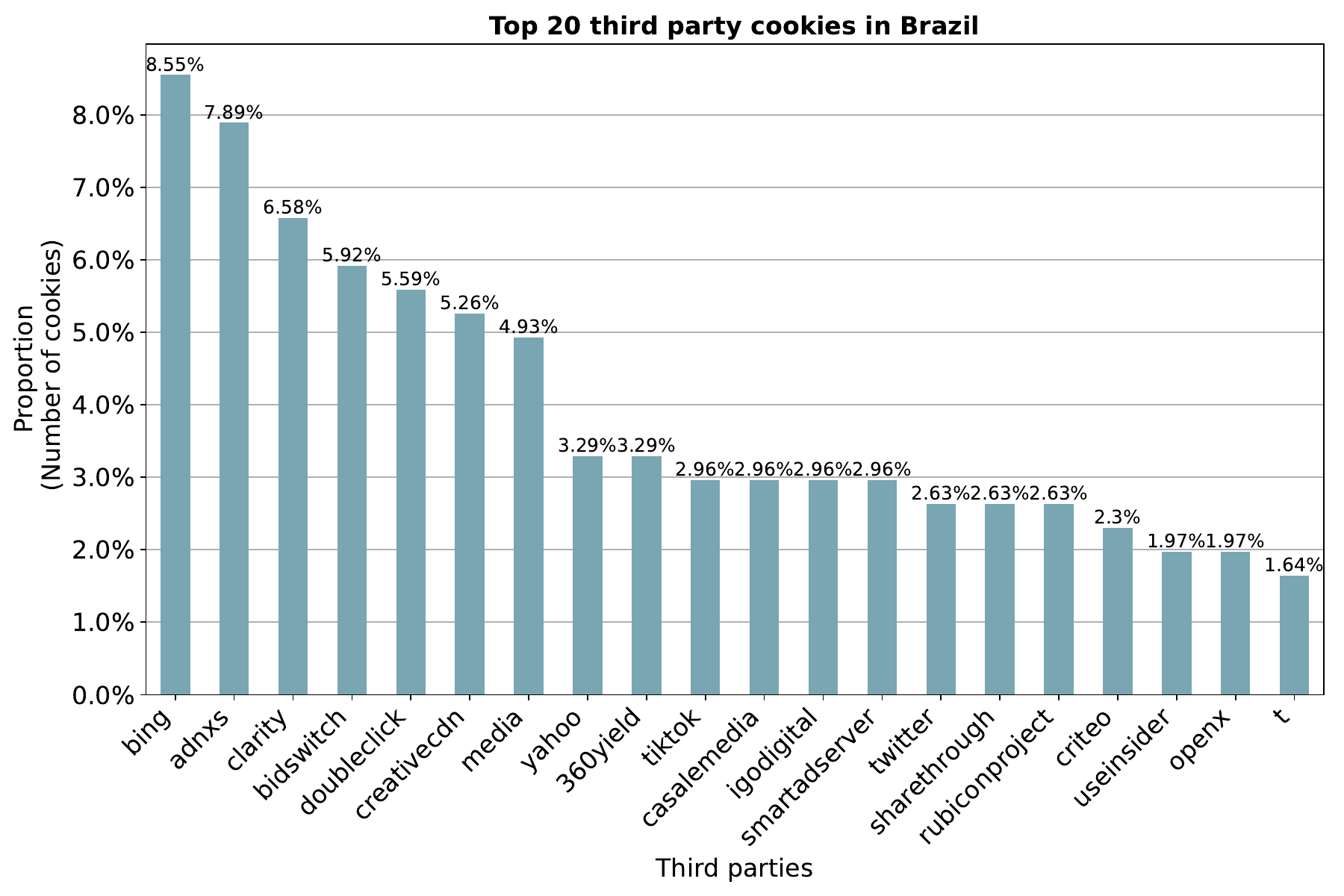}
  \caption{Third-party cookies distribution in Brazil. On a global scale, the prominent trackers `bing' and `adnxs' are top two trackers.}
  \Description{Third-party cookies distribution in Brazil.}
  \label{fig:fig 15}
\end{figure}

\begin{figure}[H]
  \centering
  \includegraphics[width=\linewidth]{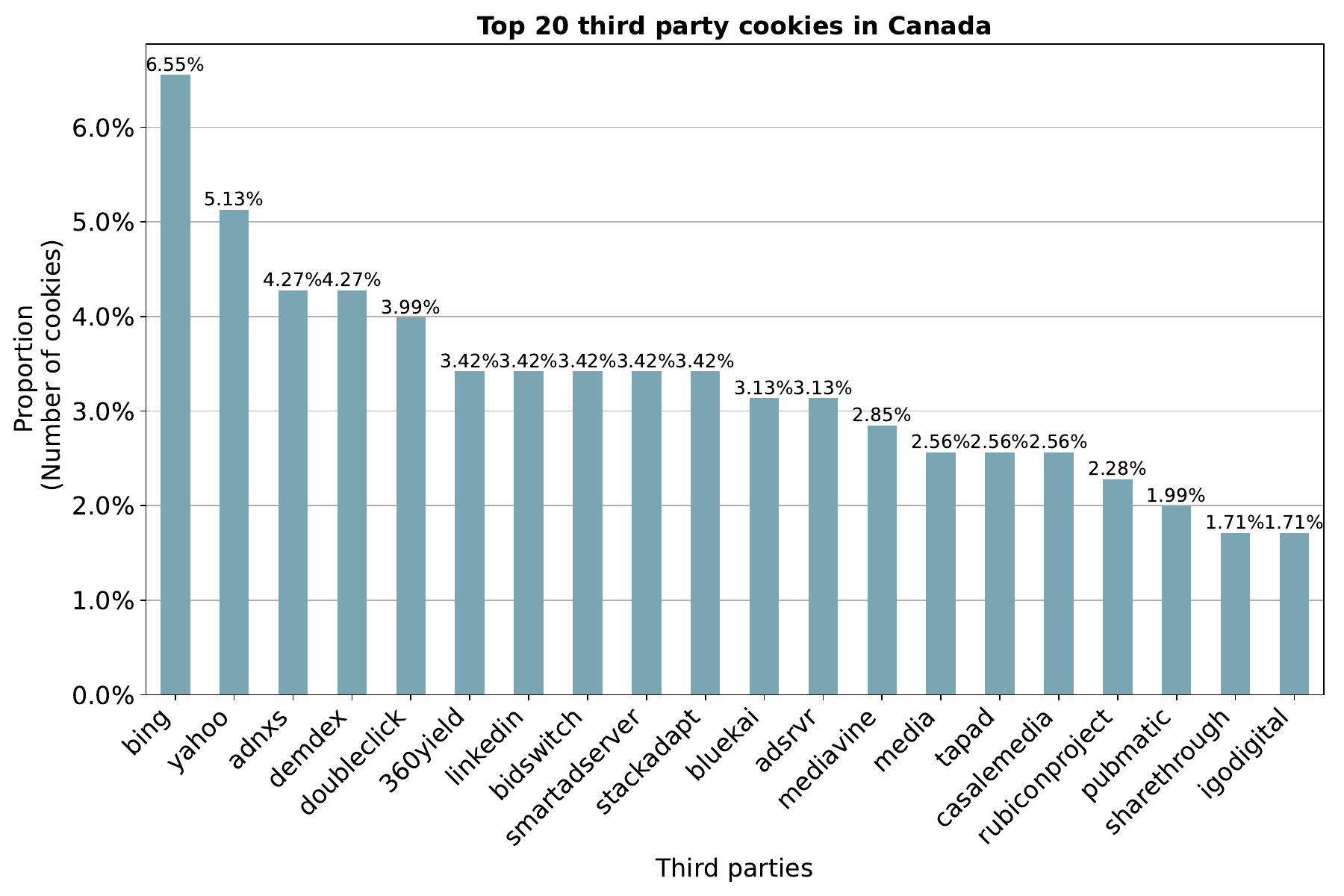}
 \caption{Third-party cookies distribution in Canada. `Bing' and `Yahoo' are the top two trackers for this country with 6\% and 5\%  cookies available respectively.}
  \Description{Third-party cookies distribution in Canada.}
  \label{fig:fig 16}
\end{figure}

\begin{figure}[H]
  \centering
  \includegraphics[width=\linewidth]{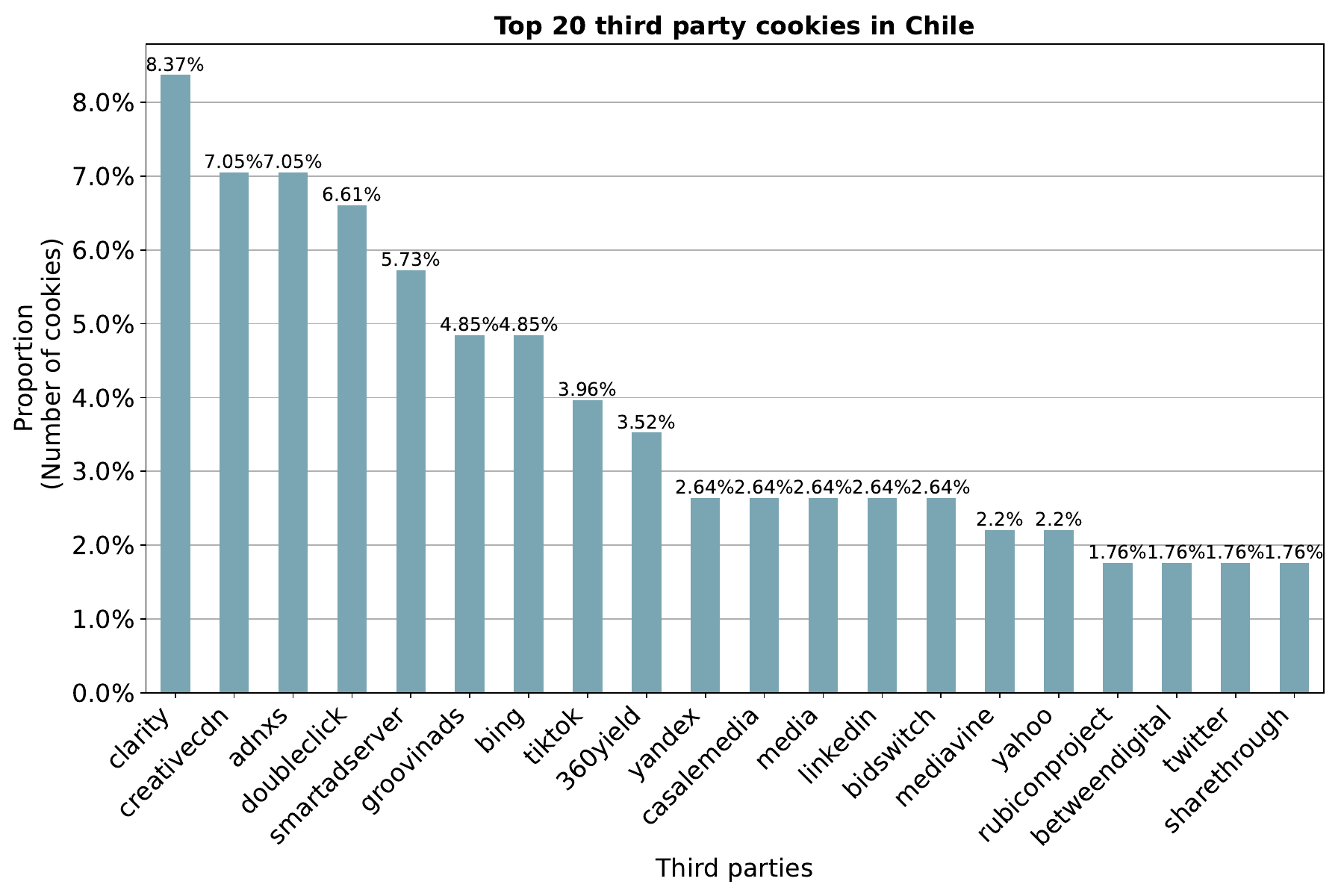}
  \caption{Third-party distribution in Chile. `Clarity' and `creativecdn' are top two trackers, positioning `adnxs' in the third position.}
  \Description{Third-party cookies distribution in China.}
  \label{fig:fig 17}
\end{figure}

\begin{figure}[H]
  \centering
  \includegraphics[width=\linewidth]{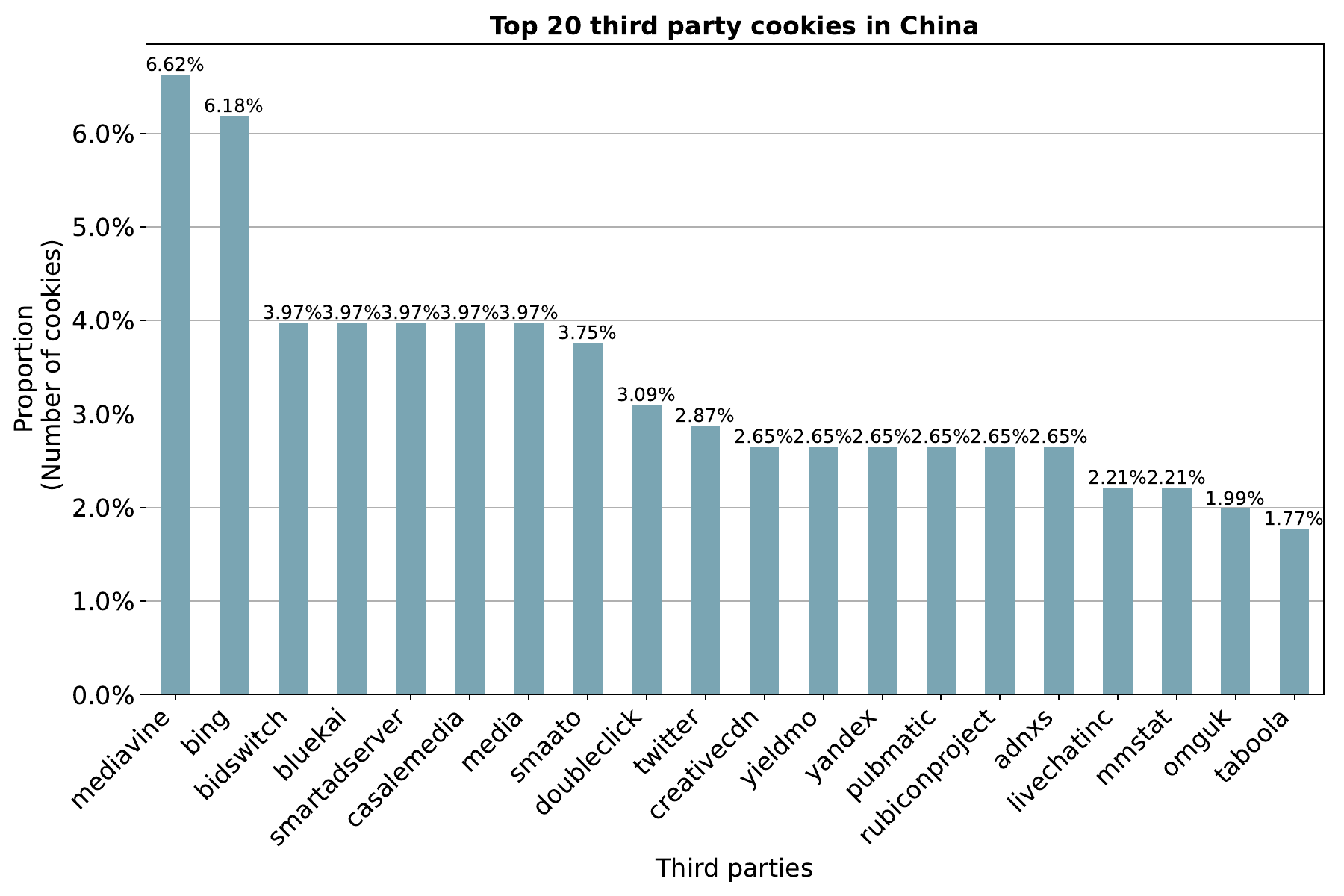}
   \caption{Third-party distribution in China. For this country, `Mediavine' is positioned at the top, with `Bing' in the second position.}
  \Description{Third-party cookies distribution in China.}
  \label{fig:fig 18}
\end{figure}

\begin{figure}[H]
  \centering
  \includegraphics[width=\linewidth]{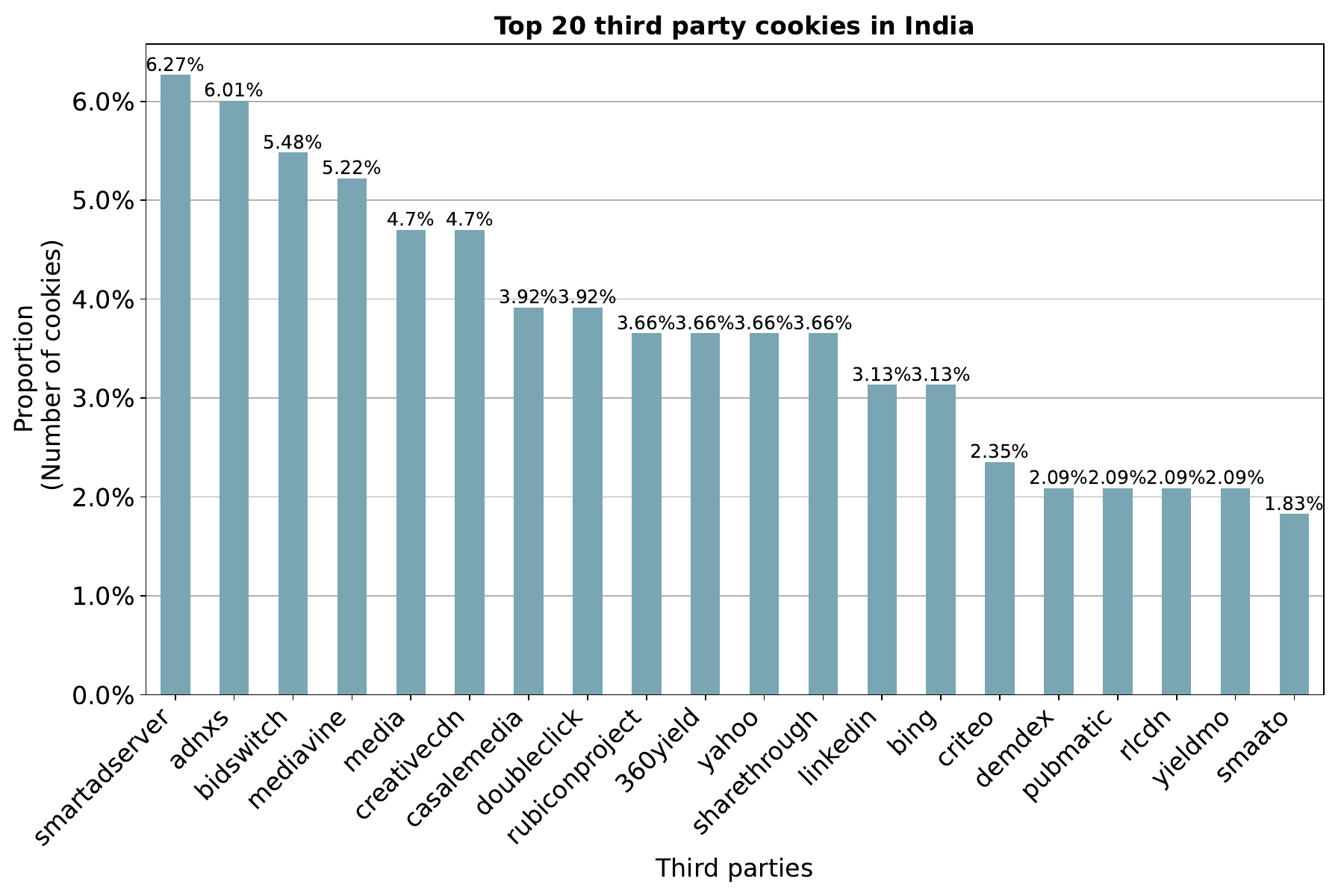}
  \caption{Third-party distribution in India. `smartadserver' and `adnxs' are among the top third-party cookies with 6\% both. The concerning issue is the dangerous, `adnxs' prevalent tracker in this country.}
  \Description{Third-party cookies distribution in India.}
  \label{fig:fig 19}
\end{figure}

\begin{figure}[H]
  \centering
  \includegraphics[width=\linewidth]{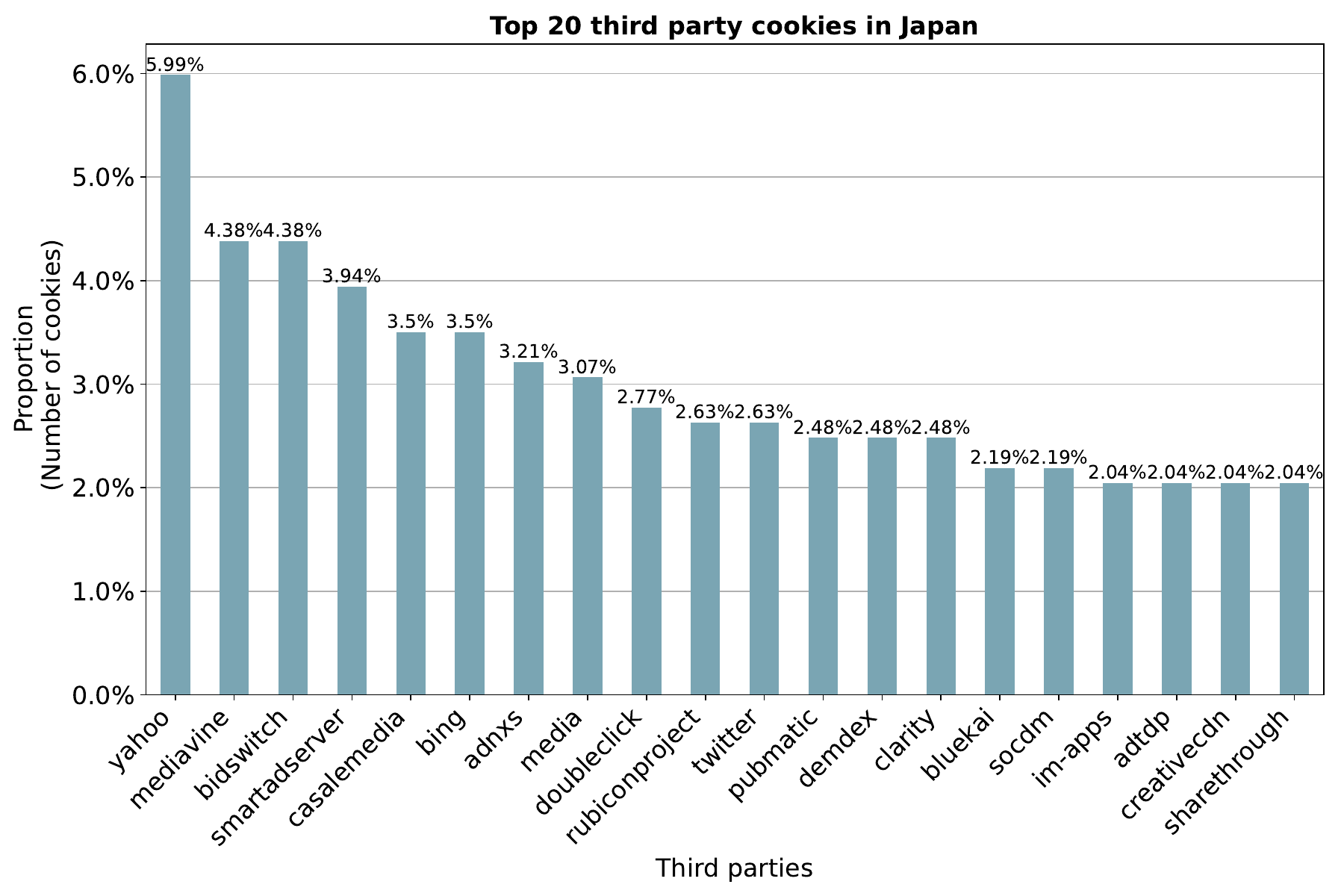}
  \caption{Third-party distribution in Japan. `Yahoo' and `mediavine' are among the top third-party cookies with 6\% and 4\% respectively.}
  \Description{Third-party cookies distribution in Japan.}
  \label{fig:fig 20}
\end{figure}

\begin{figure}[H]
  \centering
  \includegraphics[width=\linewidth]{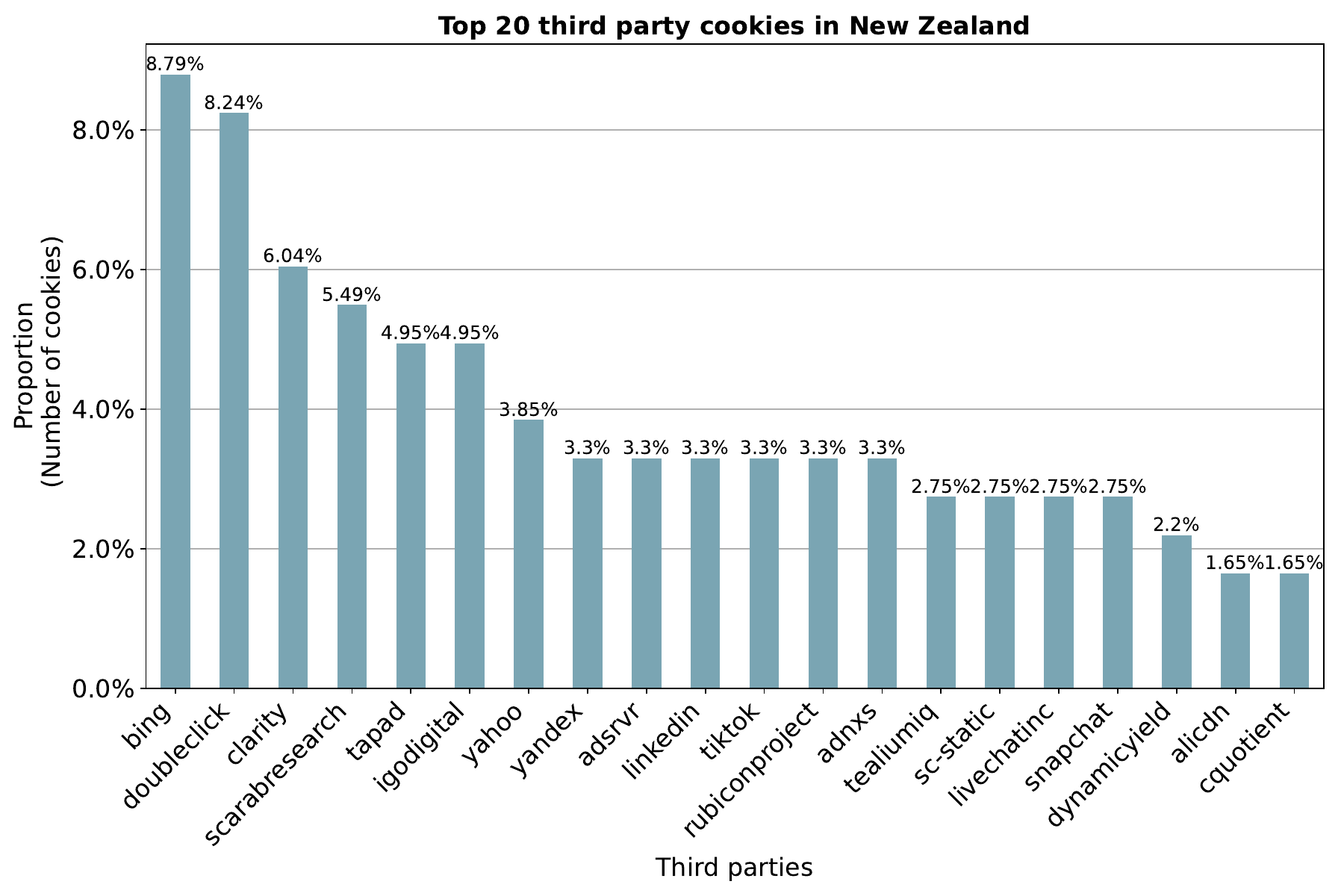}
  \caption{Third-party distribution in New Zealand.`Bing' and `doubleclick' are among the top third-party cookies, with 8\% both. Social networking cookies like TikTok and Snapchat are also prevalent in this country.}
  \Description{Third-party cookies distribution in New Zealand.}
  \label{fig:fig 21}
\end{figure}

\begin{figure}[H]
  \centering
  \includegraphics[width=\linewidth]{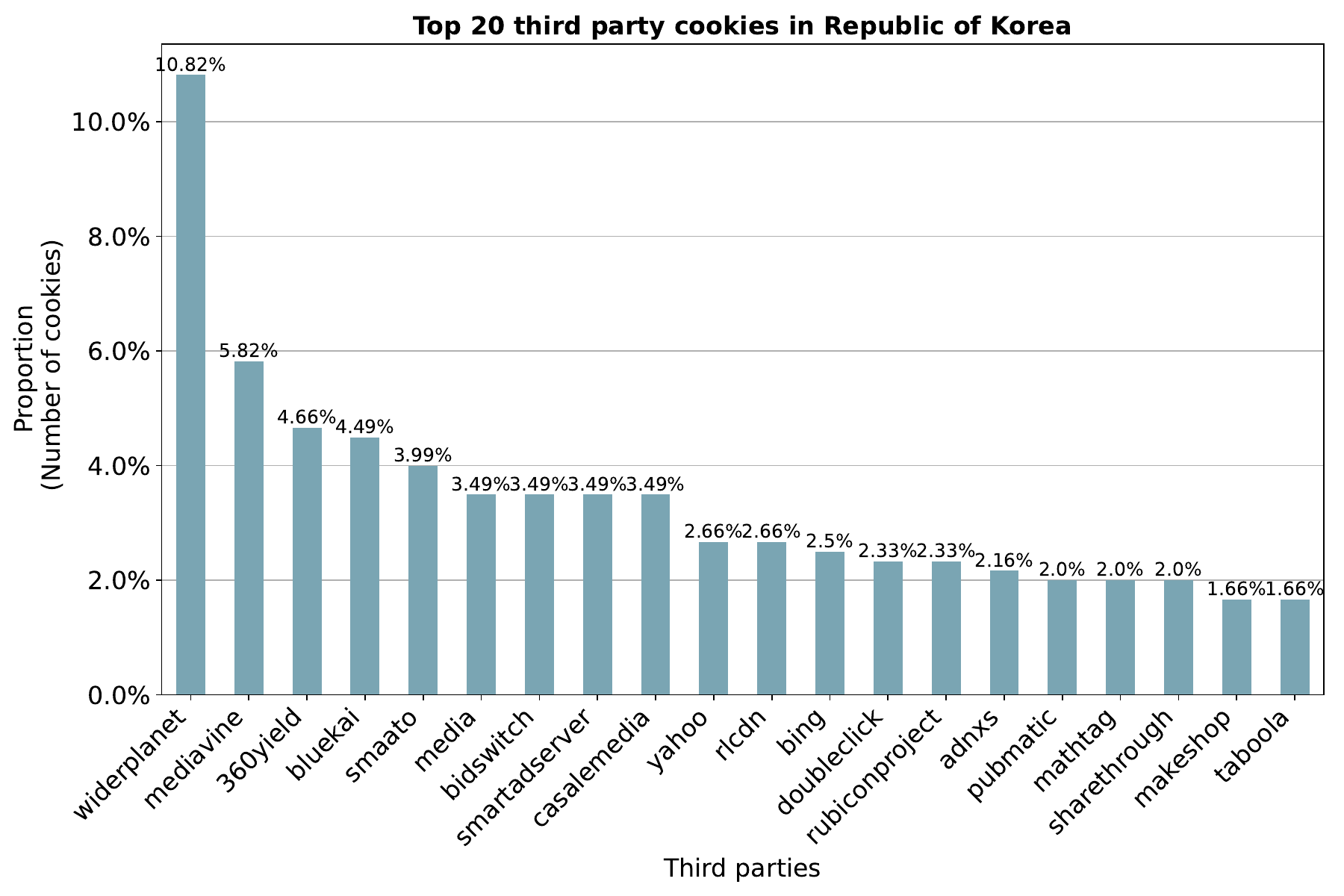}
  \caption{Third-party cookies distribution in the Republic of Korea. The most concerning issue for this country is the top tracker `widerplanet', a targeted advertising service based on big data. The cookie prevalence is as much high in comparison to other countries, having 10\% of cookies in 20 e-commerce platforms.}
  \Description{Third-party cookies distribution in the Republic of Korea.}
  \label{fig:fig 22}
\end{figure}

\begin{figure}[H]
  \centering
  \includegraphics[width=\linewidth]{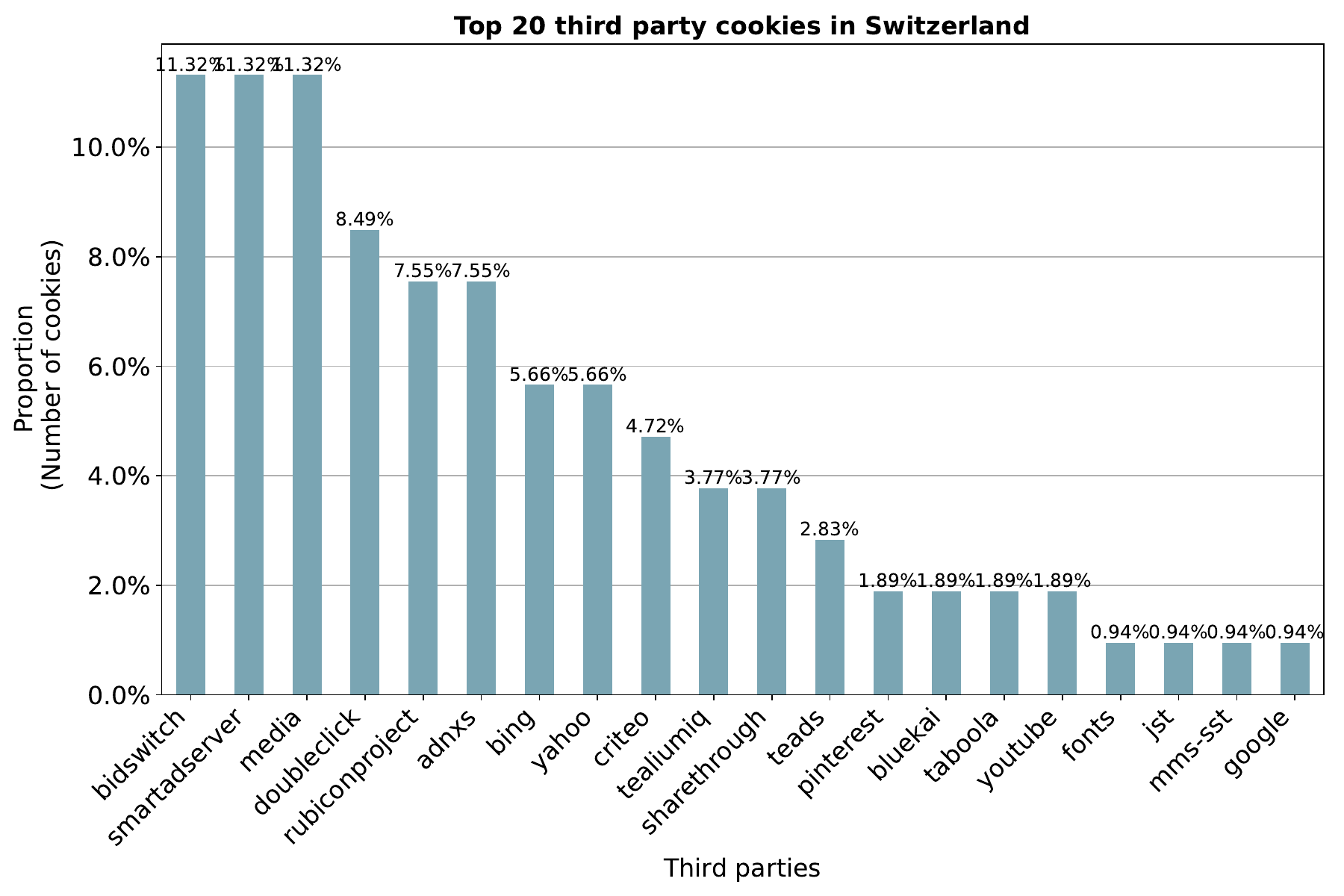}
  \caption{Third-party cookies distribution in Switzerland. `bidswitch' and `smartadsrver' are among the top third-party tracker cookies.} 
  \Description{Third-party cookies distribution in Switzerland.}
  \label{fig:fig 23}
\end{figure}

\begin{figure}[H]
  \centering
  \includegraphics[width=\linewidth]{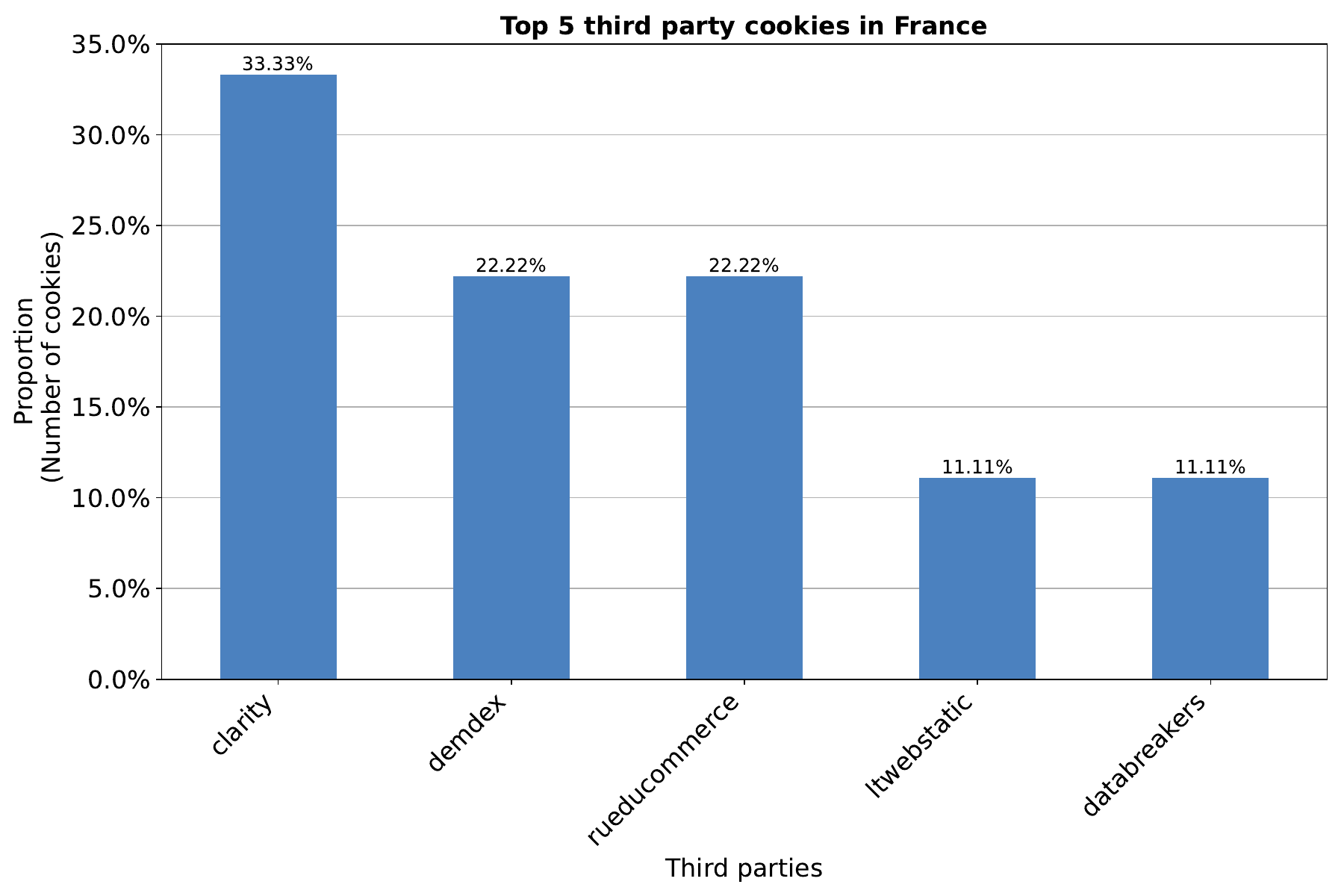}
  \caption{Third-party cookies distribution in France. Due to strict GRPR regulations, third-party cookies are less in comparison to other countries.}
  
  \Description{Third-party cookies distribution in France.}
  \label{fig:fig 24}
\end{figure}

\begin{figure}[H]
  \centering
  \includegraphics[width=\linewidth]{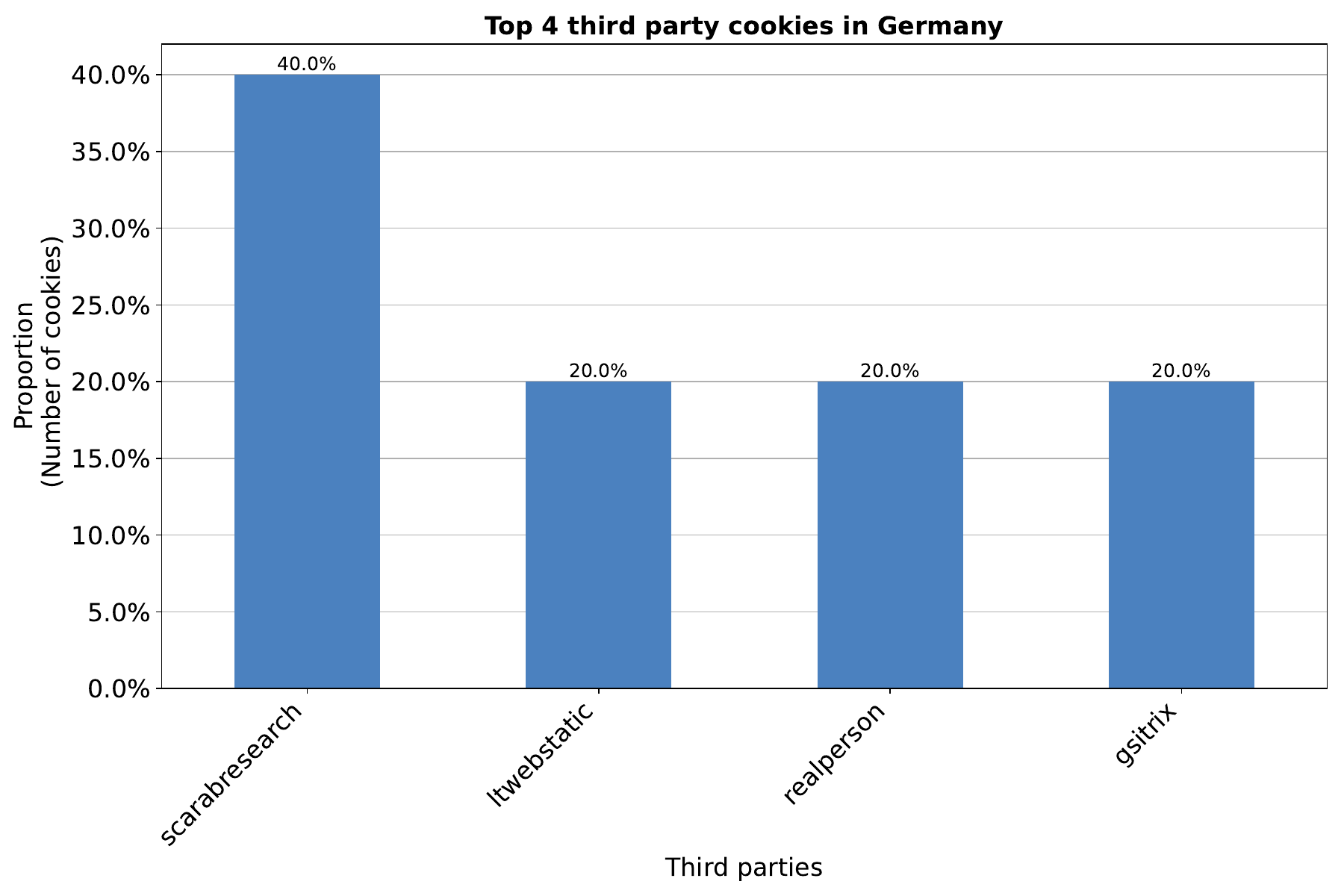}
  \caption{Third-party cookies distribution in Germany. Due to GDPR rule, third-party prevalence is less in comparison to less-stricter rule countries.}
  \Description{Third-party cookies distribution in Germany.}
  \label{fig:fig 25}
\end{figure}

\begin{figure}[H]
  \centering
  \includegraphics[width=\linewidth]{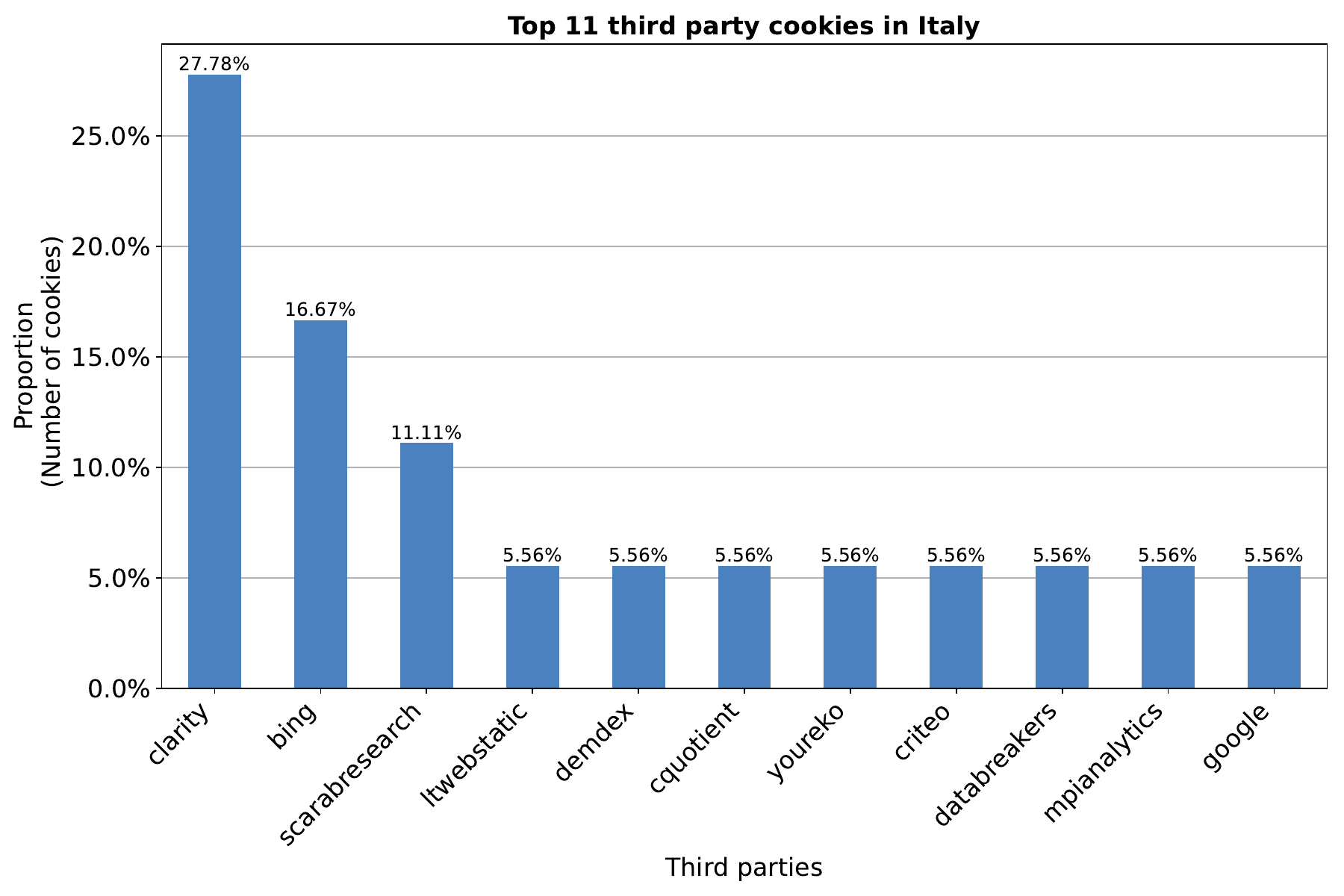}
  \caption{Third-party cookies distribution in Italy. Remarkably, this illustration depicts the diversity in the implementation of GDPR rules across different countries.}
  \Description{Third-party cookies distribution in Italy.}
  \label{fig:fig 26}
\end{figure}

\begin{figure}[H]
  \centering
  \includegraphics[width=\linewidth]{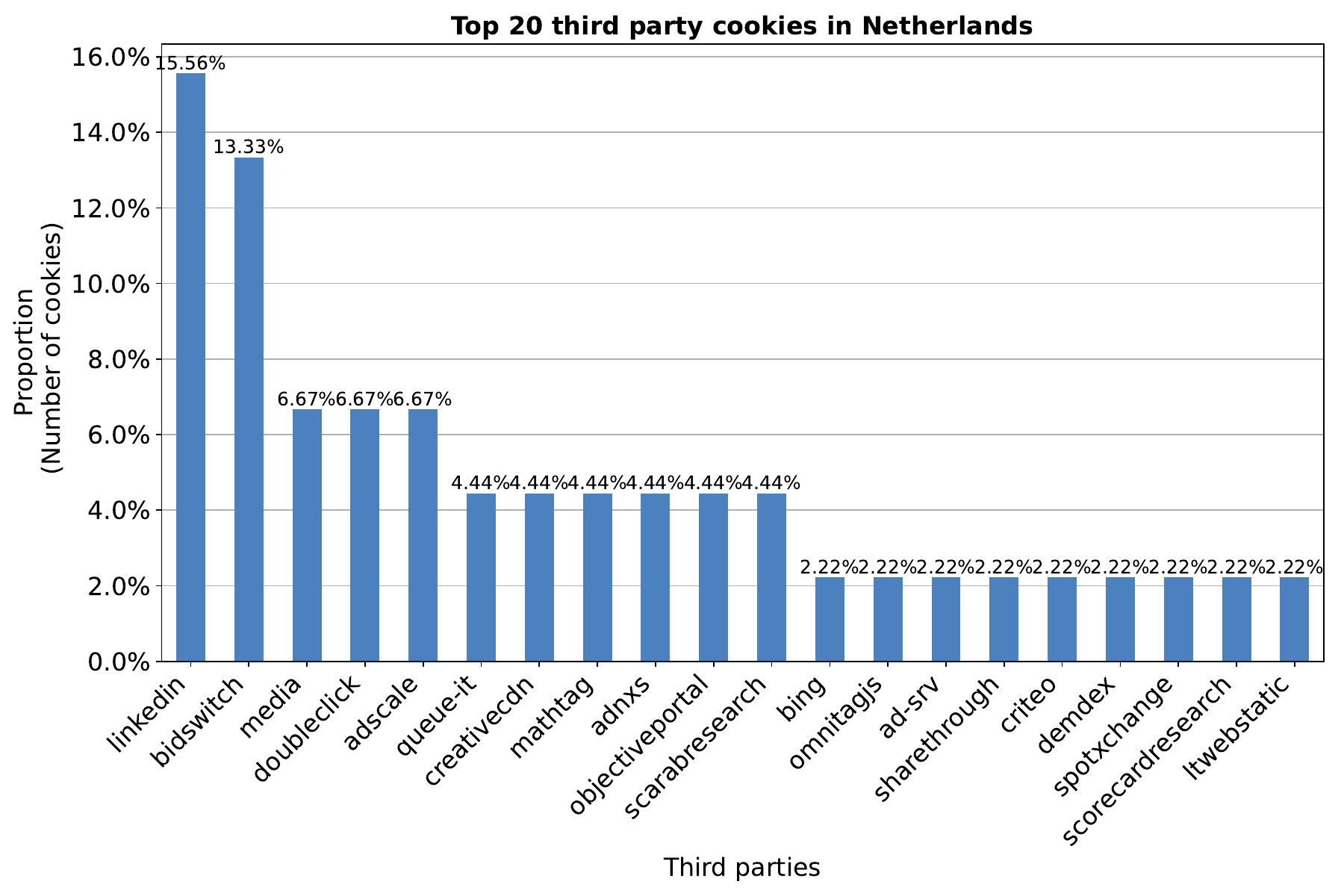}
  \caption{Tracker cookies distribution in the Netherlands. 'LinkedIn' and `bidswitch' are the top two trackers.}
  \Description{Third-party cookies distribution in the Netherlands.}
  \label{fig:fig 27}
\end{figure}

\begin{figure}[H]
  \centering
  \includegraphics[width=\linewidth]{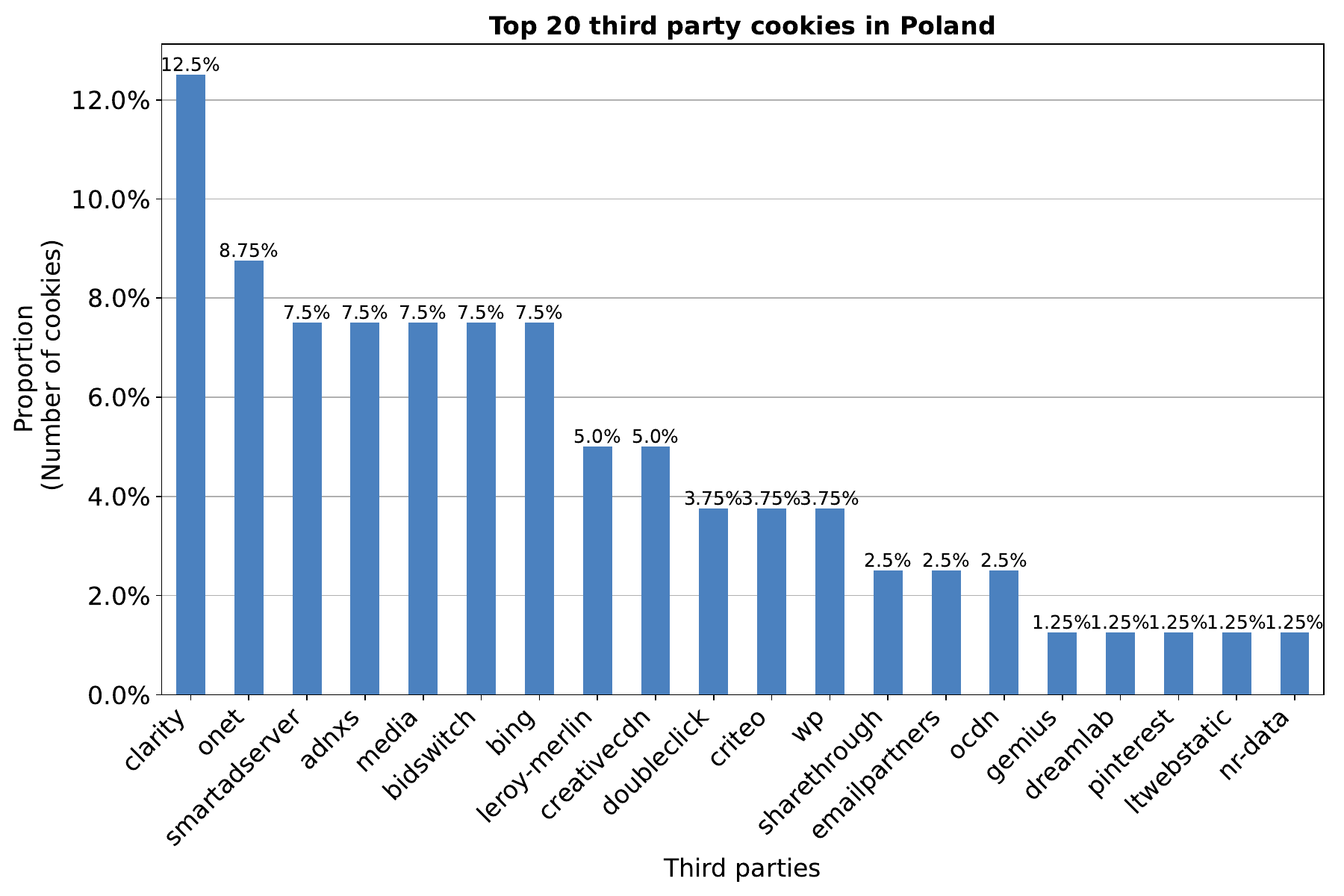}
  \caption{Tracker cookies distribution in Poland. `Clarity' and `onnet' are the top two trackers, with 12\% and 8\%  cookies.}
  \Description{Third-party cookies distribution in Poland.}
  \label{fig:fig 28}
\end{figure}

\begin{figure}[H]
  \centering
  \includegraphics[width=\linewidth]{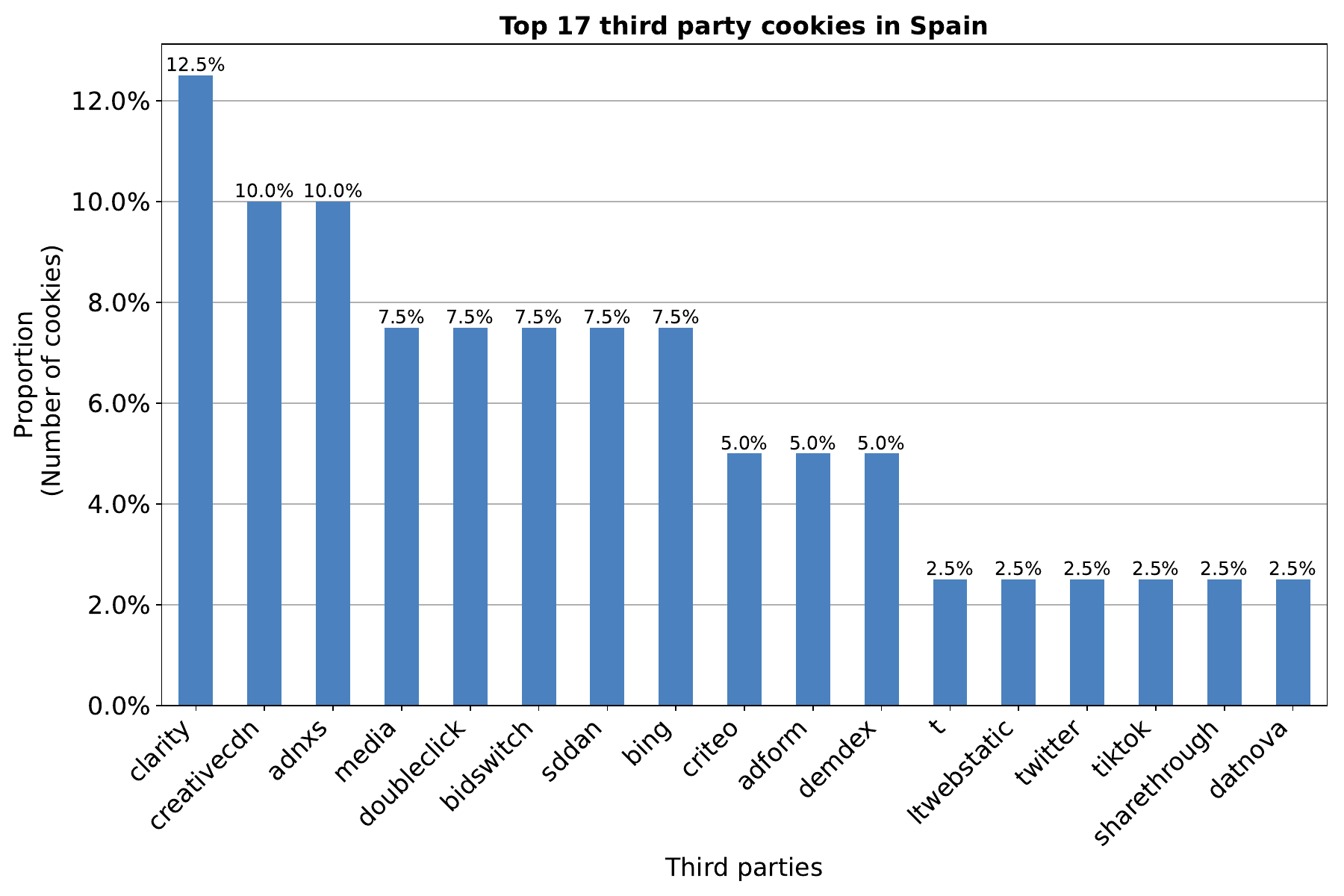}
  \caption{Third-party cookies distribution in Spain. `Clarity' and `creativecdn' are the top two trackers.} 
  \label{fig:fig 29}
\end{figure}

\begin{figure}[H]
  \centering
  \includegraphics[width=\linewidth]{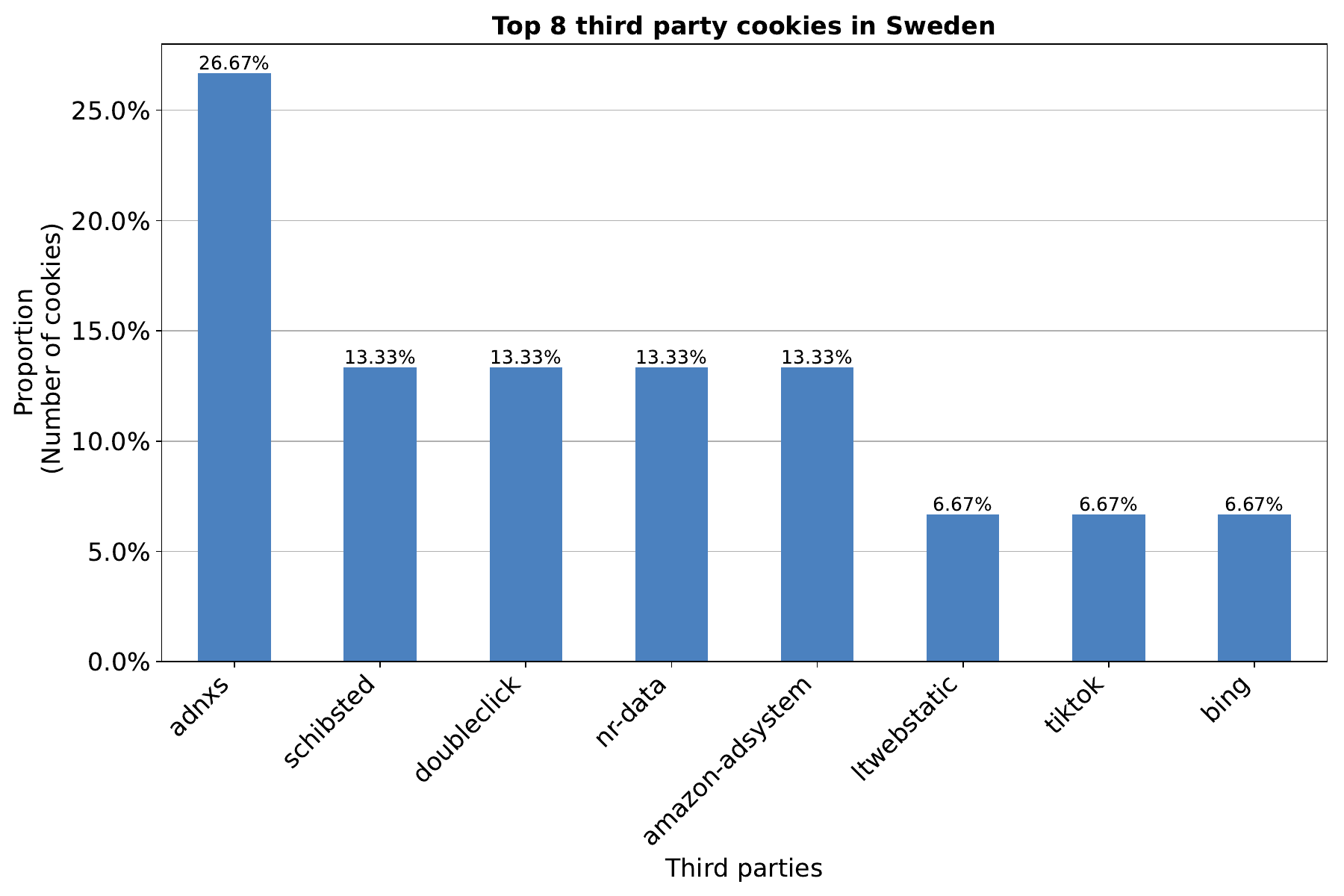}
  \caption{Third-party cookies distribution in Sweden.} 
  \Description{Third-party cookies distribution in Sweden.}
  \label{fig:fig 30}
\end{figure}

\begin{figure}[H]
  \centering
  \includegraphics[width=\linewidth]{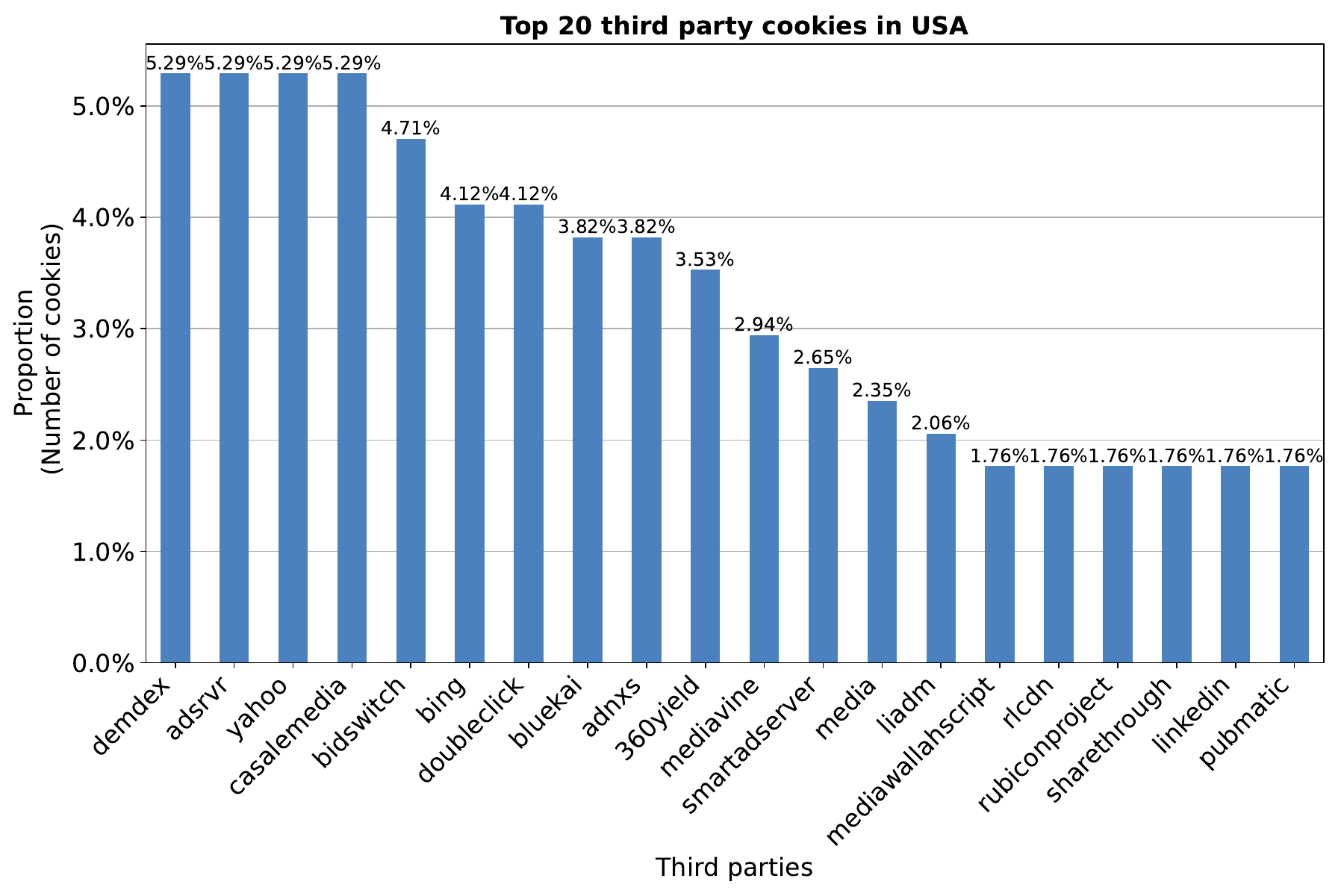}
  \caption{Third-party cookies distribution in the USA. `Demdex' and `adsrvr' are among the top third-party tracker.}
  \Description{Third-party cookies distribution in the USA}
  \label{fig:fig 31}
\end{figure}

This section illustrates the occurrence of third-party entities across 18 different countries (see Figures~\ref{fig:fig 14} through~\ref{fig:fig 31}).
Analyzing these visual representations provides insight into the predominant third-party entities in specific nations. Implementing certain limitations on these entities might contribute to safeguarding user privacy to a certain degree. Also, we include the proportion of the SameSite attribute showing distribution for three flags (See Figure~\ref{fig:samesite}).

\section{Instances of Masquerading cookies}
We provide instances of masquerading cookies in our experiment (see Table~\ref{tab:my-table}).
\begin{table*}[ht]
\caption{Instances of enumerated masquerading results}
\label{tab:my-table}
\resizebox{\textwidth}{!}{%
\begin{tabular}{@{}cl@{}}
\toprule
\textbf{Masquerading cookie} & \multicolumn{1}{c}{\textbf{First-party domain}}                             \\ \midrule
AliExpress.com               & AliExpress.nl(Netherlands)                                                  \\
Amazon.com                   & Amazon.fr(France), and Amazon.de(Germany)                                   \\
Asos.com                     & Asos.fr(France)                                                             \\
Boohoo.com                   & nz.Boohoo(New Zealand)                                                      \\
eBay.com                     & eBay.ch(Switzerland), eBay.es(Spain), and eBay.pl(Poland)                   \\
H\&M.com                     & H\&M.de(Germany), H\&M.nl(Netherlands), H\&M.in(India), and H\&M.se(Sweden) \\
Shein.com & Shein.de(Germany), Shein.it(Italy), Shein.nl(Netherlands), Shein.pl(Poland), Shein.es(Spain), Shein.fr(France), and Shein.au(Austrailia) \\
Zara.com  & Zara.de (Germany), Zara.it (Italy), Zara.nl (Netherlands), Zara.es(Spain), Zara.se(Sweden), and Zara.fr(France)                          \\ \bottomrule
\end{tabular}%
}
\end{table*}




\begin{figure}[H]
 \centering
 \includegraphics[width=\linewidth]{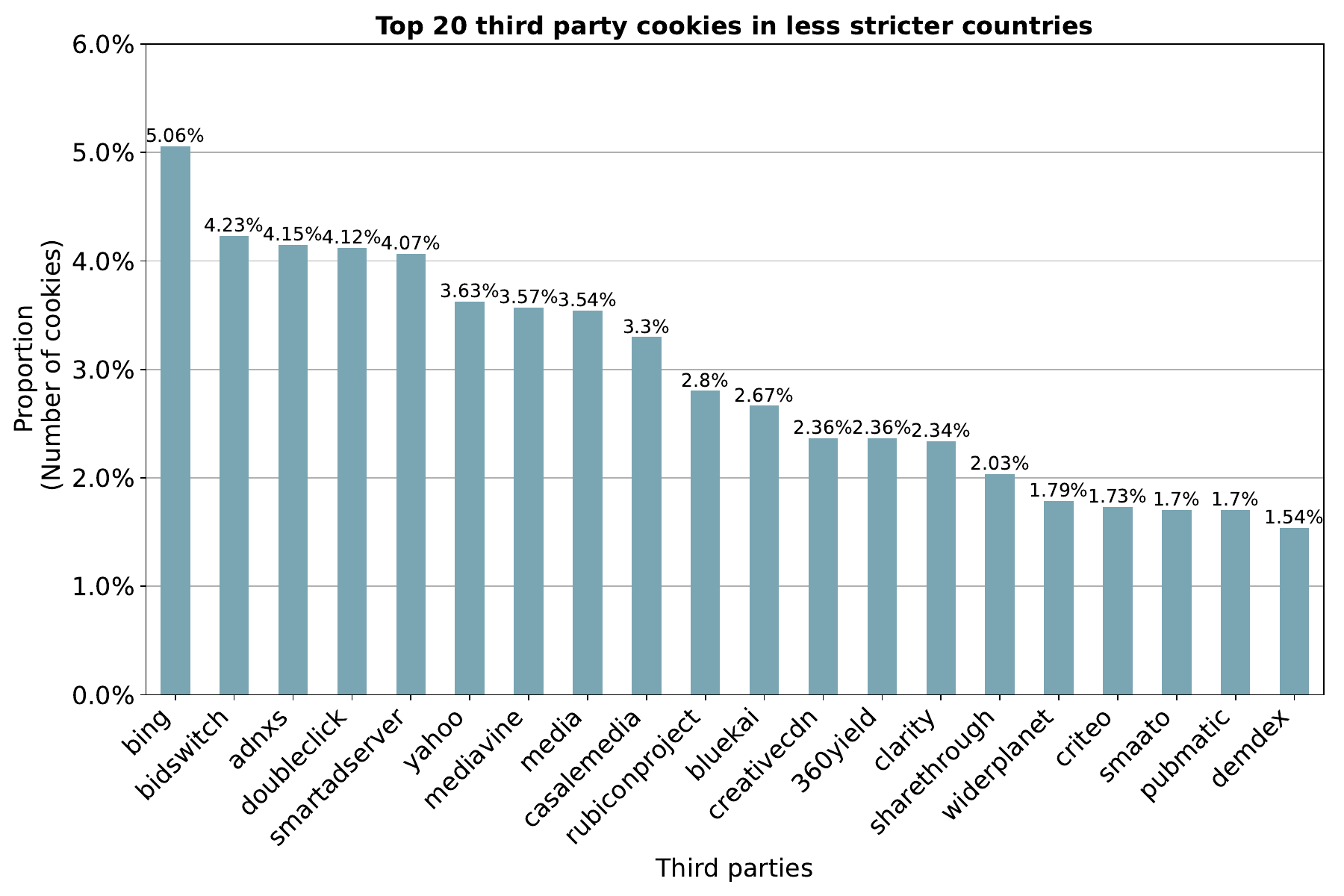}
 \caption{Top 20 third-party cookies from GDPR-like countries.}
 \Description{Top twenty third-party organizations}
 \label{fig:fig4}
\end{figure}

\begin{figure}[H]
 \centering
 \includegraphics[width=\linewidth]{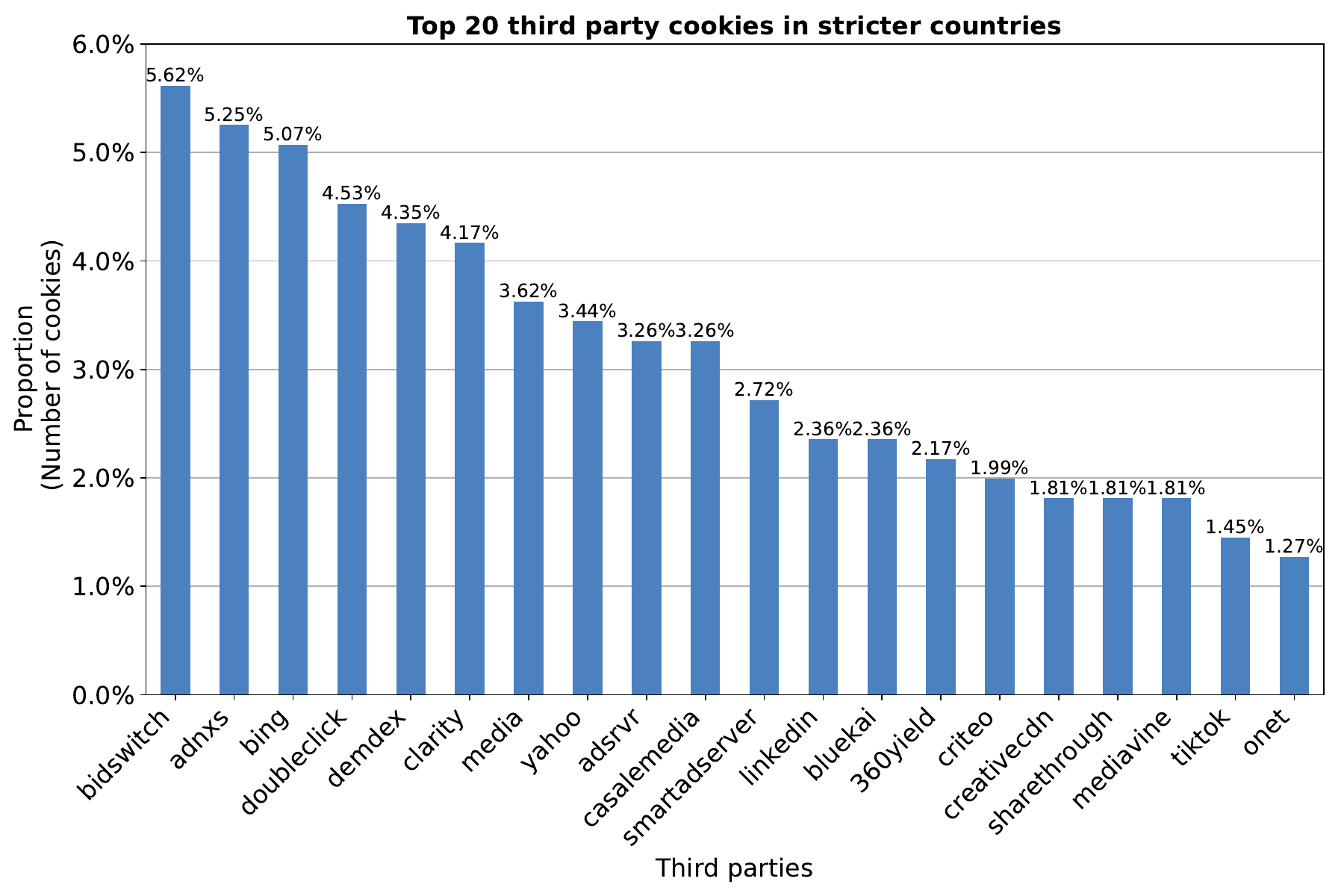}
 \caption{Top 20 third-party cookies from GDPR/CCPA countries. bidswitch is the top occuring third-party with 5.62\%.}
 \Description{Top 20 third-party appearances from GDPR/CCPA countries. bidswitch is the top occuring third-party with 5.62\%}
  \label{fig:fig5}
\end{figure}

\begin{figure}[H]
 \centering
 \includegraphics[width=\linewidth]{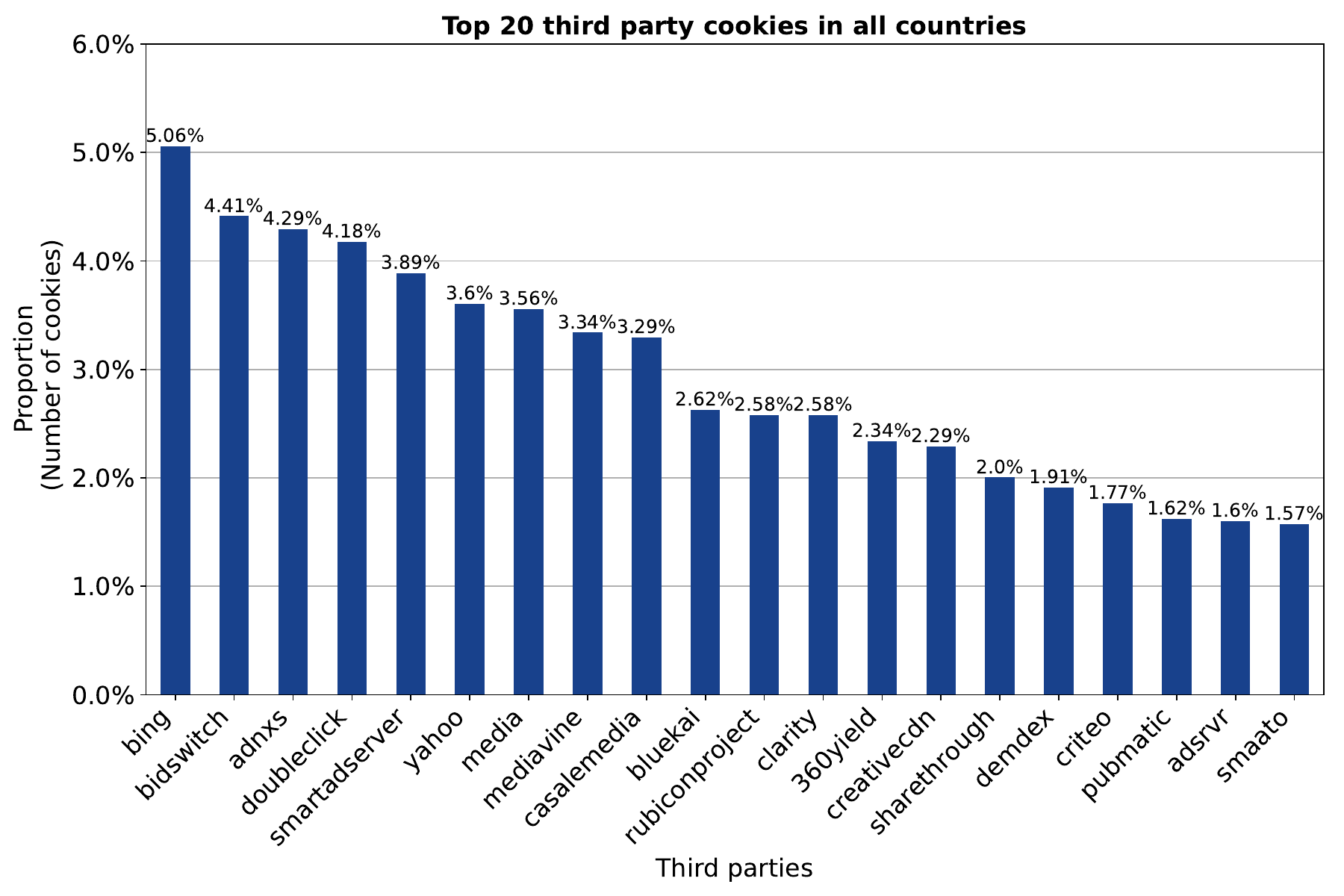}
 \caption{Top 20 third-party appearances from all 18 countries in our dataset. Bing appears as the top occurring third-party with 5.06\% while bidswitch appeares with 4.41\%.}
 \Description{Top 20 third-party appearances from all 18 countries in our dataset. Bing appears as the top occurring third-party with 5.06\% while bidswitch appeares the second most with 4.41\%}
  \label{fig:Thirdpartyallcountry}
\end{figure}
\begin{figure}[h]
 \centering
 \includegraphics[width=\linewidth]{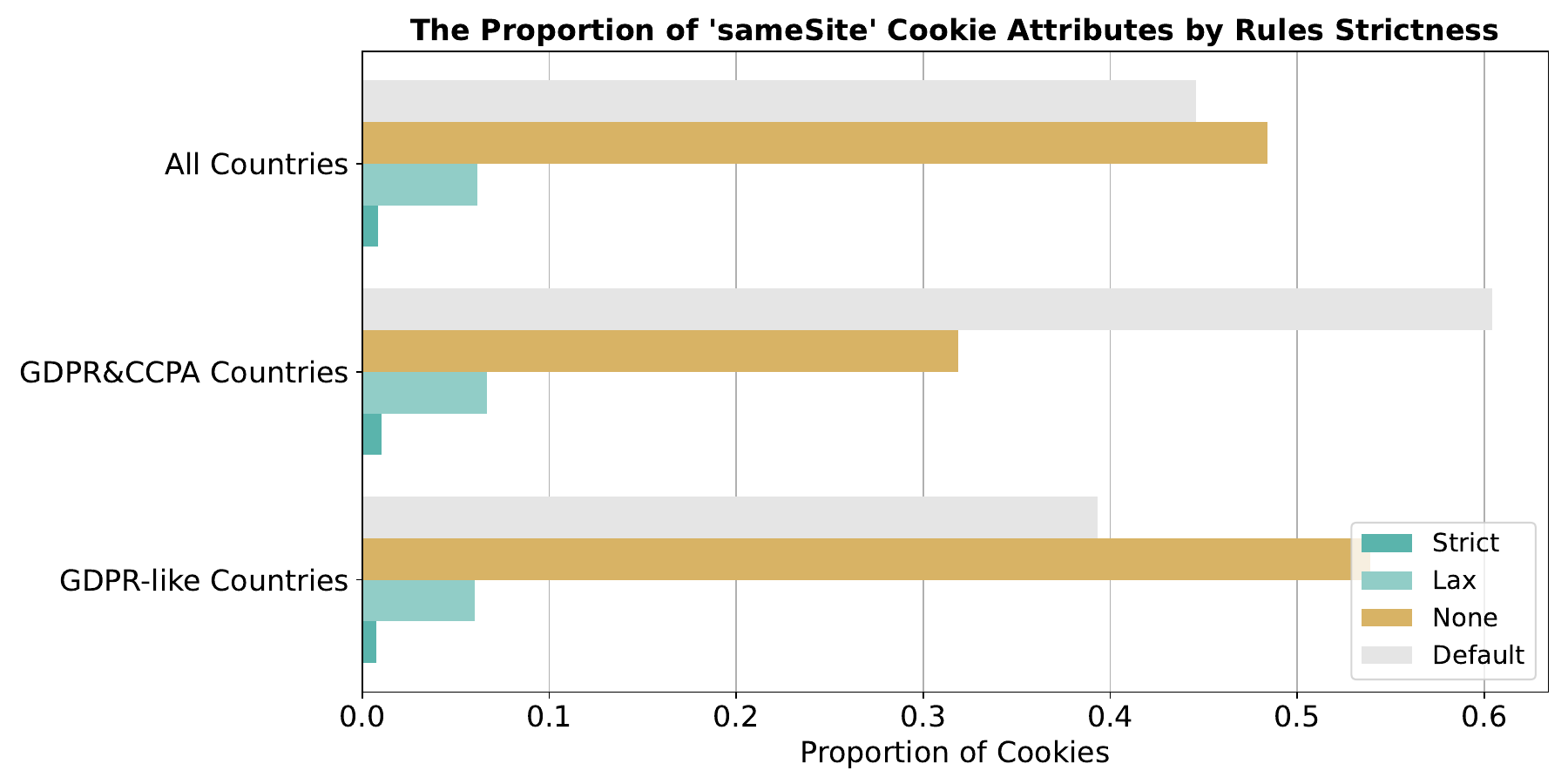}
 \caption{sameSite attribute flag proportion from all 18 countries. Figure shows very less adoption of strict flag and high adoption of flag by GDPR/CCPA country. GDPR-like countries adoptd none flag as maximum.}
 \Description{Samesite attribute for all countries}
  \label{fig:samesite}
\end{figure}